# The Evolution of Machine Learning Potentials for Molecules, Reactions and Materials

Junfan Xia[1,2], Yaolong Zhang[3] and Bin Jiang[1,2,4*]

1. State Key Laboratory of Precision and Intelligent Chemistry, University of Science and Technology of China, Hefei, Anhui 230026, China
2. School of Chemistry and Materials Science, Department of Chemical Physics, University of Science and Technology of China, Hefei, Anhui 230026, China
3. Department of Chemistry and Chemical Biology, Center for Computational Chemistry, University of New Mexico, Albuquerque, New Mexico 87131, USA
4. Hefei National Laboratory, University of Science and Technology of China, Hefei, 230088, China.

*: corresponding author: bjiangch@ustc.edu.cn

## Abstract

Recent years have witnessed the fast development of machine learning potentials (MLPs) and their widespread applications in chemistry, physics, and material science. By fitting discrete *ab initio* data faithfully to continuous and symmetry-preserving mathematical forms, MLPs have enabled accurate and efficient atomistic simulations in a large scale from first principles. In this review, we provide an overview of the evolution of MLPs in the past two decades and focus on the state-of-the-art MLPs proposed in the last a few years for molecules, reactions, and materials. We discuss some representative applications of MLPs and the trend of developing universal potentials across a variety of systems. Finally, we outline a list of open challenges and opportunities in the development and applications of MLPs.

## I. INTRODUCTION

As a foundation of quantum chemistry, the Born-Oppenheimer (BO) approximation allows for separating the nuclear and electronic motion and solving the electronic Schrödinger equation parametrically dependent on nuclear coordinates. Corresponding (adiabatic) eigen-energies at different nuclear geometries constitute a continuous hypersurface, namely the potential energy surface (PES), manifesting as a multi-dimensional function of nuclear coordinates. Based on the PES, the equations of nuclear motion can be solved either classically or quantum mechanically[1]. Thus, the PES plays a central role in achieving an atomic level of understanding of structures from small molecules to bulk materials, and dynamical processes like phase transitions and chemical reactions.

Mathematically, constructing the PES amounts to fitting or interpolating discrete potential energy data points at selected nuclear arrangements, thereby enabling energy predictions for any nuclear configuration. While conceptually straightforward, developing accurate PESs has been a longstanding challenge. Early methods of PES construction have been very dependent on the system dimensionality. For example, for condensed-phase materials or biomolecules, physically-motivated empirical "force fields" or "interatomic potentials" are often useful, which avoid the unaffordable costs for solving electronic structures.[2-5] While efficient and relatively transferable to systems of different sizes with similar elemental compositions when properly parameterized, these classical force fields often have limited expressiveness and struggle to accurately describe complex many-body interactions and bond formation/breaking processes, apart from a few exceptions[6]. On the other hand, PESs of small molecules and reactions have been largely developed by fitting or interpolating *ab initio* data with more sophisticated mathematic expressions[7-9]. This goes under the broad heading of the machine learning (ML) term — regression[10]. Such PESs provide high flexibility, catering to the demands of studying molecular spectroscopy and reaction dynamics for a specific system, but neither transferable to others nor scalable to larger systems. For medium-sized systems up to a few hundreds of atoms, alternatively, it is possible to compute the required potential energy and atomic forces on-the-fly as nuclear geometry varies, avoiding the difficulties of explicitly constructing analytical PESs. This direct dynamics or *ab initio* molecular dynamics (AIMD) method[11, 12] is typically coupled with computationally inexpensive electronic structure methods, such as density functional theory (DFT). Even so, AIMD simulations are limited to short timescales and difficult to study rare events such as chemical reactions. Moreover, unless the wave function is highly localized[13, 14], on-the-fly quantum dynamics simulations are generally unfeasible.

Modern ML techniques, such as neural networks (NNs)[15], Kernel-based methods[16], and many others[10], have recently become powerful tools across scientific fields, due to their ability to effectively capture complex relationships in high-dimensional spaces[17]. One particularly fruitful application of ML in the physical sciences is the ML representation of multi-dimensional PESs for diverse systems. Alongside taking advantage of ML ability to accurately approximate real-valued functions, considerable attention has been paid on ensuring that the PES remains invariant under symmetry operations, notably permutation of like atoms. The corresponding symmetry-adapted ML potentials (MLPs) achieve remarkable flexibility and generalizability in mapping structures to corresponding energies and forces. MLPs have been maturely applied to a wide range of scenarios, such as chemical reactions[18-24], material properties[25-29], heterogenous systems[30-33], excited states[34-38], spectroscopy[39-47], and more.

Since this is a rather active and fast-growing field with widespread applications, it is almost impossible to cover all aspects about MLPs in a single review. There have existed many excellent reviews on this topic from different points of view[19, 37, 48-58]. The present review aims to trace the evolution of MLP methodologies, from early models to current state-of-the-art approaches, and to outline future trends in the field. Note that basic concepts of ML algorithms have been well described in literature and will not be repeated here.

The review is organized as follows. The next section provides a concise overview of the early stages of MLPs, highlighting two distinct development strategies: one based on the global representation, and the other on an atomic energy decomposition with predefined local or iteratively learnable descriptors. Section III reviews recent developments of MLPs following both strategies for describing more complex systems and relevant long-range effects. Section IV presents some representative applications and discusses the trend towards developing universal potentials across the periodic table. Finally, Section V concludes by highlighting the challenges and opportunities in this rapidly evolving field.

## II. Brief Overview of Machine Learning Potentials in Early Stages

The application of ML methods to represent PESs has a long history. The earliest attempts to use NNs to fit PESs emerged in the late 1990s for molecule-surface systems and small molecules.[59, 60] Preliminary efforts were also made in using NNs to fit empirical parameters of force fields for condensed-phase systems.[61] However, limitations in high-quality data and computational resources restricted these early studies to proof-of-concept demonstrations, just showcasing the potential of NNs as a superior nonlinear fitting tool compared to empirical functions for representing PESs. Importantly, the proper symmetry adaption of the PES, particularly its invariance with respect to the permutation of identical atoms, was not yet systematically addressed. While some ideas for symmetrizing either the input or hidden layer of NNs had been proposed in specific cases[62, 63], a general solution was lacking.

Investigations of MLPs became more prevalent in the 2000s in small systems. For example, Raff and coworkers published a series of studies on NN potentials (NNPs) for polyatomic molecules and reactions integrated with novelty sampling of *ab initio* data points.[64-66] Manzhos and Carrington made seminal contributions in developing *ab initio* NNPs for small molecules reaching spectroscopic accuracy based on a sum-of-products form[67] or an *N*-mode expansion form[68], where individual NNs were used to fit low-mode component functions[69, 70]. Alternatively, in a similar many-body (MB) expansion, Malshe *et al.* directly learned low body-order terms instead of component functions by NNs[71]. Unfortunately, permutation symmetry was not satisfied in these representations. Symmetry-adapted coordinates started to be used in NNPs of molecule-surface systems by Lorenz *et al.* [72] in 2004, which were later generalized by Reuter and coworkers using Fourier expansions of atomic Cartesian coordinates on different crystal facets, though restricted to diatomic molecules on rigid surfaces[73].

It is worth mentioning that, in this period, NNs were not the dominant way of constructing PESs. For molecular systems, many other non-NN interpolation/fitting methods have been developed in parallel, such as the modified Shepard interpolation (MSI)[9], reproducing kernel Hilbert space (RKHS) interpolation[74], interpolation moving least squares (IMLS)[75], and permutationally invariant polynomial (PIP) fitting[76]. These methods can be actually regarded as generalized ML regression methods as none of them rely on a physically-derived functional form[77]. For example, the RKHS method[74] is a kernel-based interpolation method by solving a linear inversion problem, which is closely related to Gaussian process regression (GPR)[16] and provides an automatic estimate of the prediction error. Furthermore, key concepts like symmetry adaptation and data sampling, initially developed alongside these non-NN methods, have been later widely recognized in the ML community. For example, Collins and coworkers[9] proposed a trajectory-based data sampling strategy coupled with their MSI method to iteratively select new data points which are most distant from existing ones and least confident in the PES prediction, which allows one to gradually refine the PES[78]. This was perhaps the earliest scheme of active learning (AL), in the language of ML, and many similar algorithms are now extensively used in the automatic construction of modern MLPs. Additionally, the PIP method proposed by Bowman and coworkers[79] is a linear regression model based on a set of symmetrized basis functions, preserving the permutational invariance of the PES. Importantly, PIPs can be systematically generated by the combinations of the functions of internuclear distances[80], providing a general way for symmetrizing the input of other ML methods. However, these NN or non-NN methods express the total energy by a single ML model as a function of the entire structure, which was typically limited to low-dimensional systems with a few atoms. This type of MLPs is hereafter referred to as global descriptor-based models.

In the same period, developing MLPs for high-dimensional and periodic systems remained very challenging. In such cases, it turns out that, empirical force fields, such as the embedded atom method (EAM)[3], have identified a naturally more scalable representation of the PES. Specifically, the total energy is decomposed into the sum of atomic contributions,

$$E = \sum_{i}^{N} E_i(\mathbf{r}_{ij}, \mathbf{r}_{ik}, \cdots) \tag{1}$$

and each atomic energy depends on the positions of neighbor atoms relative to the central atom within a truncated local environment. Unfortunately, the atomic energy in empirical force fields was often expressed as empirical functions of bond distances (and angles) with very limited expressiveness. Smith and coworkers[81] realized the advantage of this local representation and employed individual atomic NNs to represent atomic energies. However, they did not address how to design the NN input vector in a general and symmetry-preserving manner.

A remarkable breakthrough was achieved by Behler and Parrinello (BP) in 2007, who first proposed to use a fixed-length array of many-body functions as the NN input vector to describe the atomic structure in general and an element-wise NN architecture to ensure that the sum of atomic energies is permutationally invariant[82]. In this BPNN approach, atom-centered symmetry functions (ACSFs)[83] were designed as local descriptors. ACSFs are calculated by the sum of functions involving the central atom and all neighbors within the local environment and naturally invariant upon permutation of any like neighbors. This elegant construction not only fulfills the symmetry requirements but also ensures that the computational cost of the PES scales linearly with respect to the total number of atoms. In addition, it can describe both molecular and periodic systems of variable sizes, provided that their atomic environments are well learned during the training process. This local representation was later coupled with the GPR method by Csányi and coworkers to develop the Gaussian Approximation Potential (GAP)[84], another pioneering local descriptor-based MLP model. As a non-parametric ML method, the GAP predictions of atomic energies are directly correlated with the similarity of atomic structures.

Since then, especially in the past ten years or so, modern MLPs have experienced rapid growth. On the one hand, PIPs[79] or more concise fundamental invariants (FIs)[85] have been coupled with NNs or GPR, enabling the construction of highly-accurate and permutationally-invariant PESs of small to medium-sized molecules in the gas phase and on metal surfaces. Global descriptor-based GPR models, like gradient domain machine learning (GDML)[86] and pKREG[87] models, have shown promise in describing relatively large molecules with a small amount of data. These active global descriptor-based MLP models are schematically illustrated in Figure 1. On the other hand, a great deal of local descriptor-based MLPs have been developed. Many of them modified the original ACSFs for better representability to atomic structures or with lower costs, while some others provide more systematic and efficient ways of generating atom-centered many-body descriptors up to an arbitrary body order. More recently, the advent of message passing neural networks (MPNNs) [88, 89] has retained the atomic energy representation while moving beyond the conventional idea of purely local descriptors. MPNNs enable learnable atomic descriptors that incorporate non-local effects, leading to a surge of innovative models with continually improved accuracy. Figure 2 illustrates the framework of these atomistic MLP models and presents a timeline of the emergence of some representative models to date. We shall next discuss these important advancements along both routes for developing MLPs.

## III. Recent Developments of MLPs

### A. Global Descriptor-based Models

To ensure the translational and rotational invariance of the PES, interatomic distances, though redundant, are a convenient choice as primitive structural descriptors in a global PES representation. However, the permutational invariance makes the situation more complicated. In practice, one can represent the potential energy using a multinomial expansion of interatomic distances (often their functions) raised to various powers, with coefficients determined by linear regression. However, it is crucial to apply all permutation operations within the molecular permutation group to these polynomials, ensuring that the resulting permutationally invariant polynomials (PIPs) have identical coefficients[80], formally expressed as follows,

$$E = \sum_{l_{ij}=0, l_{ik}=0,\cdots}^{l} c_{l_{ij}, l_{ik},\cdots} \hat{S}\left[\prod_{i<j}^{N} y_{ij}^{l_{ij}}(r_{ij})\right], \qquad l = \sum_{i<j}^{N} l_{ij} \tag{2}$$

where $\hat{S}$ is a symmetrization operator and $l$ is the sum of the powers of exponential functions of interatomic distances ( $y_{ij}^{l_{ij}}(r_{ij}) = \exp(-\alpha r_{ij})$ ensuring the correct decay behavior in the asymptote). By construction, this PIP method is simple and clear. Bowman and coworkers derived PIPs up to a certain order systematically through low-order primary invariants and secondary invariants[76], making it a very efficient method for constructing PESs of small molecules and reactions involving many like atoms, as have been discussed in several excellent reviews[76, 90]. A disadvantage of this method is that the total number of PIPs (and the maximum order) needed to form a complete basis set fulfilling permutation invariance increases combinatorically as the number of allowed permutations. Consequently, the generation of PIPs for large polyatomics became very time-consuming. Using a library of PIPs that was developed in the middle 2000s, PIP potentials have been developed up to 10 atoms. More recently, due to a fragmented and pruned approach, PIP potentials have made possible for molecular systems with up to 21 atoms[91-95]. To further extend its applicability, the Paesani and Bowman groups expressed the total energy in the MB expansion form and used the PIP method to represent each expanded MB term. This MB-PIP scheme allowed them to successfully construct a series of highly accurate potentials of bulk water[96-98] and generic molecules[99] . Comparison with other MLP methods, PIP was also found to be more data-efficient and much faster in small molecular systems[100-102]. Very recently, Both the MB-PIP and fragmented PIP approaches have been extended to construct the PESs of $C_{14}H_{30}$, pushing the limit of PIP potentials to a molecular system with 44 atoms[103]. It is worth noting that open-source software packages of the PIP method, PESPIP[104] and MOLPIPx[105], have been available, enabling the efficient computation of energy gradients and even higher-order derivatives in an end-ot-end manner[105].

Given the pros of PIPs and the high representability of NNs, Jiang, Li and Guo and coworkers proposed a permutation invariant polynomial neural network (PIP-NN) method[106-108]. In this method, PIPs are used as descriptors of NNs to ensure the overall symmetry of the system, while NNs are used to achieve high accuracy of the PES. The PIP-NN method have been applied to molecules and reactions with up to 8 atoms in gas phase[109] and 6 atoms on a rigid surface[110], as have been summarized in several reviews[111, 112]. Pruned PIPs have been also used in the PIP-NN method in some non-reactive systems[113]. Because of the high expressive power of NN, the PIP-NN method usually leads to smaller fitting errors than the PIP method with sufficient data[111, 112].

Nevertheless, PIPs are non-unique and often redundant, thus not the optimal global descriptors for NNs. For a given molecular permutation group, there exist a group of FIs serving as the generators for PIPs up to an arbitrary order[85]. This inspires Fu, Zhang, and coworkers to propose the fundamental invariant neural network (FI-NN) method[114]. Importantly, FIs are linearly independent of each other and using FIs as the input of NN minimizes the size of global descriptors with correct symmetry. Even so, the generation of FIs remains to suffer from unfavorable combinatorial scaling with respect to the number of identical atoms in molecules, and the corresponding FIs become very lengthy. Chen *et al.* developed an efficient parallel algorithm that enables the generation of a manageable number of FI terms (typically 500~800) by a reasonable truncation of the highest order to five or six, providing to adequate accuracy of the PES[115]. By doing this, they have successfully applied the FI-NN method to develop global PESs for bimolecular reactions involving up to 15 atoms[116], as have discussed in their recent review[24].

It should be emphasized that, in theory, both PIPs and FIs need to reach a certain order and include necessary terms in each order to rigorously guarantee the permutation invariance of the system. We found examples in both molecular and molecule-surface systems where incomplete terms of PIPs cannot distinguish two seemingly different configurations. This deficiency leads to much increased fitting errors[107, 108]. More generally, incomplete descriptors—for example, those lacking high-order many-body correlations—can cause distinct configurations to become indistinguishable in descriptor space, thereby significantly limiting the representational power of the ML model. This issue was deeply discussed by Ceriotti and coworkers in relation to atomic structures.[117-119]

Although methods such as PIP-NN[106-108] and FI-NN[114] are of high accuracy, they usually rely on tens of thousands of data points for a given polyatomic system. Kernel-based regression methods, such as GPR[16], have the potential to work well for small datasets. For example, Kamath *et al.* compared GPR with NN in representing the PES of formaldehyde with the same number of data points[120]. They found the GPR-based PES is more accurate when using only 625 points, while both PESs become comparably accurate when using 2500 points. Krems and coworkers showed that GPR models with composite kernels instead of a single kernel function can be iteratively improved by increasing the complexity of kernels and can be used for physical extrapolation of PES[121]. This technique allowed them to construct the global GPR-based PES with low-energy data points for the protonated imidazole dimer, a molecular system with 19 atoms[122]. Symmetry-adaption has also been considered in GPR models. For example, PIPs have been also coupled with GPR yielding the PIP-GPR method, which was found to achieve comparable accuracy as the PIP method in selected testcases[123]. GPR and kernel ridge regression (KRR) methods are closely related. Dral and coworkers proposed a global descriptor-based KRR model using relative-to-equilibrium (RE) distances (*i.e.* $r_{ij}^{eq} / r_{ij}$) and a Gaussian kernel, KREG for short[124]. Its permutationally invariant version, pKREG[87], incorporates permutational invariance of kernel functions and analytical gradients[125]. This significantly improves the model accuracy for small datasets in developing global PESs for molecules with up to 22 atoms, involving multiple conformers[125].

A special kernel-based approach using global descriptors is the gradient domain machine learning (GDML) method[22] developed by Tkatchenko, Müller and coworkers. Unlike most other MLPs, interestingly, GDML just fits forces and directly predicts them rather than first fitting energy and then calculating forces as the negative gradient of energy with respect to the nuclear coordinates. To ensure energy conservation, the key is to define the kernel $\mathbf{K}(\mathbf{x},\mathbf{x}') = \nabla_{\mathbf{x}} K_E(\mathbf{x},\mathbf{x}') \nabla_{\mathbf{x}'}^{\mathrm{T}}$, which models the force vector ($\mathbf{F}$) as a transformation of an unknown energy ($E$),

$$\mathbf{F} = -\nabla E \approx \mathrm{GPR}\left[-\nabla \mu_E(\mathbf{x}), \mathbf{K}(\mathbf{x},\mathbf{x}')\right] \tag{3}$$

Here, $\mathbf{x}$ is the global structural descriptor vector consisting of the inverse of all pairwise distances, $\mu_E(\mathbf{x})$ and $K_E(\mathbf{x},\mathbf{x}')$ are the prior mean and prior covariance functions that define the latent energy-based GPR, respectively. The original GDML showed exceptional performance against small datasets of molecules, but was not symmetry-preserving. Later, a symmetric GDML (sGDML)[86] model was modified by incorporating spatial and temporal physical symmetries in an automatic data-driven way, which enabled force continuity and efficient molecular dynamics (MD) simulations of flexible molecules with up to a few dozens of atoms with high accuracy. Very recently, the same group further extended this GDML approach to handle even more flexible molecules containing more than one hundred atoms[126]. This globally iterative sGDML scheme[126] reduces the computational complexity, lowering the original quadratic scaling to a linear scaling with respect to the system size. On the other hand, a Bravais-inspired GDML (BIGDML) method[127] was proposed to account for periodicity and the invariance with respect to the full symmetry group of the parent lattice. It was not only applicable to two-dimensional and bulk materials, but also adsorbates on surfaces. These advances have made GDML-based approaches promising in dealing with more extended systems, while retaining the advantages of the global representation, *e.g.* for including long-range interactions.

**B. Local Descriptor-based Models**

Despite the advances mentioned above, global descriptor-based MLPs by construction suffer from a rapid increase in complexity and computational cost as the system size grows. In addition, the descriptor size is non-uniform but system-specific, making it difficult to train an MLP with a mixed data of multiple systems and limiting the transferability of an MLP across diverse systems. By contrast, local descriptor-based models, as conceptually illustrated in Figure 2(a), can more easily handle these problems. Obviously, the choice of atomic descriptors is critical to the description of the atomic environment and thus the performance of the model. Although a variety of local descriptor-based MLP architectures have been proposed from different concepts, they can generally be categorized based on the order of interactions in their descriptors, such as two-body, three-body, or higher-order many-body interactions.

The first generation of BPNN uses two-body and three-body ACSFs as atomic descriptors.[83] The two-body symmetry function can be often written as follows,

$$G_i^2 = \sum_{j\neq i}^{N_c} \exp\left(-\alpha (r_{ij} - r_s)^2\right) f_c(r_{ij}) \tag{4}$$

where $N_c$ is the number of neighboring atoms within the cutoff radius ($r_c$), $r_{ij}$ is the interatomic distance between the central atom $i$ and neighboring atom $j$, and the radial function here is a Gaussian function in which $\alpha$ and $r_s$ represent the width and center, respectively. Other functional forms were later proposed in place of the Gaussian function, such as Bessel functions[128], polynomial functions[129], and power functions[130], which can offer faster computation or improved representability. $f_c(r_{ij})$ is a cutoff function used to smoothly truncate interactions between atoms that exceed $r_c$, often taking the form of a cosine function or its power function.[83] Some studies have employed polynomials as cutoff functions or incorporated cutoff functions into the radial functions to reduce computational costs.[129, 131]

Two-body descriptors based solely on interatomic distances are insufficient to fully differentiate atomic local environments. To improve the representability, the three-body descriptor adopted by Behler and Parrinello is a mix of the radial and angular functions, for example[83],

$$G_i^3 = \sum_{j\neq i}^{N_c} \sum_{k\neq i,j}^{N_c} \frac{1}{2^{\xi-1}} \left(1+\lambda\cos(\theta_{ijk})\right)^{\xi} \exp\left(-\alpha(r_{ij}^2 + r_{ik}^2)\right) f_c(r_{ij}) f_c(r_{ik}) \tag{5}$$

in which $\theta_{ijk}$ is the enclosed angle between neighbors $j$ and $k$ and the central atom $i$, $\lambda$ and $\xi$ define the shape of the angular function. While the three-body symmetry function is more expressive, the double loop over neighboring atoms in Eq. (5) results in a quadratic scaling with $N_c$, $\sim\mathcal{O}(N_c^2)$. Furthermore, the choice of hyper-parameters often relies on human experiences. Consequently, the BPNN method has not become popular and seldom been applied to complex systems with more than three elements in the first a few years.

More improved variants of ACSFs have been proposed in recent years. For example, Smith *et al.* first developed a ANI-1 potential[132] for a variety number of organic molecules with multiple elements by using element-specific pairwise interactions and incorporating angular shifting parameters into the ACSFs, which increases the distinguishability of the atomic descriptors. However, the number of ACSFs increases significantly with the combinations of elements. A similar effect is achieved in the weighted ACSF (wASCF) method by Gastegger *et al.* using an element-weighted summation instead of individual element-dependent descriptors for obtaining a better scaling with the increasing number of elements.[133] Imbalzano *et al.* developed several feature selection algorithms, which can choose the suitable hyperparameters of ACSFs automatically.[134] Yao *et al.* developed the TensorMol model, which adjusts symmetry function hyper-parameters along with NN parameters during the training process.[135] ACSF-like descriptors have been widely available in many NNP packages, for example, RuNNer by Belher and coworkers[136], ænet by Artrith and Urban[137], and *n2p2* by Dellago and coworkers[138], which later used low-cost polynomial radial functions for improved description of atomic environments[129]. We note that a kernel-based, Faber-Christensen-Huang-Lilienfeld (FCHL19) model[139], also utilizes ACSF-like descriptors for the similarity metric and energy predictions. These descriptors all rely on explicit expression of three-body functions.

Other three-body descriptors have been proposed in different mathematical forms or physical concepts. For example, a popular Deep Potential (DP) model, introduced by Zhang *et al.*, describes the local environment via the inner product of the generalized coordinate matrix, which is generated by multiplying atomic Cartesian coordinates with NN-based radial functions.[140] This effectively constituted implicit three-body descriptors, with parameters determined by an embedded NN, but favoring a linear-scaling computational cost ($\sim \mathcal{O}(N_c)$). The DeePMD-kit, an efficient and user-friendly open-source package based on the DP method, enables automatic construction of DP models for beginners and facilitates the widespread applications of this approach.[141]

Inspired by the EAM[3], we developed the embedded atom neural network (EANN) method based on the embedded atom density (EAD) descriptor[142]. An embedding density feature, $\rho_i$, of atom *i* can be regarded as the square of the linear combination of the Gaussian type orbitals (GTOs), $\varphi_{l_x l_y l_z}(\mathbf{r}_{ij})$, of neighboring atoms,

$$\rho_i = \sum_{l_x,l_y,l_z}^{l=l_x+l_y+l_z} \frac{l!}{l_x!l_y!l_z!} \left[ \sum_{j \neq i}^{N_c} c_j \varphi_{l_x l_y l_z}(\mathbf{r}_{ij}) \right]^2 \tag{6}$$

$$\varphi_{l_x l_y l_z}(\mathbf{r}_{ij}) = x_{ij}^{l_x} y_{ij}^{l_y} z_{ij}^{l_z} R(r_{ij}) f_c(r_{ij}) \quad (7)$$

where $l=l_x+l_y+l_z$ specifies the total angular momentum of the orbit and its components in each dimension, $\mathbf{r}_{ij} \equiv (x_{ij}, y_{ij}, z_{ij})$ is the Cartesian coordinate vector between atom $i$ and $j$, $r_{ij}$ is the interatomic distance, $R(r_{ij})$ is a Gaussian-type radial function, and $f_c(r_{ij})$ is a cutoff function. The orbital coefficient $c_j$ is an element-dependent parameter that ensures the permutation invariance of like atoms, and can be easily optimized during the training process. It should be noted that the EAD descriptor includes three-body ($l$>0) interactions implicitly, leading to a linear scaling with $N_c$, *i.e.* $\sim\mathcal{O}(N_c)$.

Local descriptors involving up to three-body correlations have generally been considered sufficient for describing atomic environments in many practical applications. Liu and coworkers included the dihedral-based four-body functions in their power-type structural descriptor[130], which has been found useful in constructing global PESs for very complicated catalyst structures and catalytic reactions[143], although the computational complexity scales cubically with $N_c$, *i.e.* $\sim\mathcal{O}(N_c^3)$. Pozdnyakov *et al.* have recently identified some counterexamples whose atomic structures cannot be well represented by three- (or even four-) body descriptors.[117] Theoretically, a complete description of the local environment with $N_c$ atoms requires a combination of many-body descriptors that are invariant to all permutations, ranging from two-body to $N_c$-body. However, this leads to a factorial increase in computational cost, making it impractical for complex molecules and condensed-phase systems with many atoms included in the neighborhood. Thus, the efficient incorporation of higher-order many-body correlations into descriptors has become crucial for improving the performance of local descriptor-based models in complex systems.

There are more systematic ways of generating many-body atomic descriptors. A pioneering descriptor of this type was the smooth overlap of atomic positions (SOAP)[144], coupled with the GAP model[84] developed by Csányi and coworkers. This descriptor encodes atomic environmental information based on the expansion of atomic neighbor density $\rho_i^a(\mathbf{r})$ for each element *a*, in a local basis of orthogonal radial function $R_n(r)$ and spherical harmonics $Y_{lm}(\hat{\mathbf{r}})$, with expansion coefficients $c_{i,nlm}$,

$$\rho_i^a(\mathbf{r}) = \sum_{nlm} c_{i,nlm}^a R_n(r) Y_{lm}(\hat{\mathbf{r}}) \tag{8}$$

Summing up the square of the expansion coefficients over the index *m* yields the rotationally invariant power spectrum, $p_{i,nn'l}^{aa'}$,

$$p_{i,nn'l}^{aa'} = \frac{1}{\sqrt{2l+1}} \sum_m (c_{i,nlm}^a)^* c_{i,n'lm}^{a'} \tag{9}$$

in which $a$ and $a'$ represent two neighbor-element channels, $n$ and $n'$ represent two radial channels and *l* represents an angular channel. One can also obtain the bispectrum analogously. In GAP, the similarity of two atomic structures is defined as the inner product of two atomic neighbor densities, characterized by the rotationally invariant SOAP kernel[144],

$$k(\rho, \rho') = \int \mathrm{d}\hat{R} \left| \int \mathrm{d}\mathbf{r} \rho(\mathbf{r}) \rho'(\hat{R}\mathbf{r}) \right|^{v} \tag{10}$$

in which $\hat{R}$ is a spatial rotation, *v* is a positive integer corresponding to the maximum of (*v*+1)th-body order. The SOAP kernel can be expanded analytically in terms of the power spectrum (*v*=2) or bispectrum (*v*=3). Similarly, Thompson *et al.* developed the spectral neighbor analysis potential (SNAP) by assuming a linear relationship between atomic energy and bispectrum components[145].

A more general approach to incorporate high-order many-body interactions is through tensor product and tensor contraction. In this context, Shapeev *et al.* developed the Moment Tensor Potential (MTP) method[146] in the Cartesian space, where the many-body correlations are expressed by contracting multiple tensors consisting of invariant polynomials and corresponding contraction coefficients are determined through linear regression to the atomic energy. This approach can be applied to include many-body interactions of arbitrary order while keeping a linear scaling, $\sim \mathcal{O}(N_c)$, of the computational complexity of each body-order term. In this regard, the EAD descriptor can be considered as a special case and generated using tensor contraction as well. The linear combination of GTOs in Eq. (6), $\boldsymbol{\psi}_{i,\boldsymbol{l}}$, can be alternatively expressed in a tensor product form,

$$\boldsymbol{\psi}_{i,\boldsymbol{l}} = \sum_{j\neq i}^{N_c} c_j R(r_{ij}) \underbrace{\mathbf{r}_{ij} \otimes \cdots \otimes \mathbf{r}_{ij}}_{l \text{ times}} \tag{11}$$

Here, the subscript $\boldsymbol{l}$ collects all the indices of the $l$-order tensor and $\mathbf{r}_{ij} \otimes \cdots \otimes \mathbf{r}_{ij}$ represents the tensor product of $\mathbf{r}_{ij}$. It should be noted that the choice of the radial function does not affect the overall evaluation. A general many-body descriptor can then be obtained by the Einstein summation over the common indices in $\boldsymbol{l}_1, \cdots, \boldsymbol{l}_v$,

$$\rho_i = \underbrace{\boldsymbol{\psi}_{i,\boldsymbol{l}_1} \cdots \boldsymbol{\psi}_{i,\boldsymbol{l}_v}}_{v \text{ tensors}} \tag{12}$$

in which $v$ tensors $\boldsymbol{\psi}_{i,\boldsymbol{l}_1}, \cdots, \boldsymbol{\psi}_{i,\boldsymbol{l}_v}$ are contracted to a zero-order tensor (scalar) $\rho_i$ with the ($v$+1)*th* body-order feature. It should be noted that the dimension of $\boldsymbol{l}_1, \cdots, \boldsymbol{l}_v$ are unnecessary to be identical, as long as they share some indices. The three-body EAD feature can then be simply obtained by $\boldsymbol{\psi}_{i,l_1}\boldsymbol{\psi}_{i,l_1}$, namely the sum of the squares of $\boldsymbol{\psi}_{i,l_1}$ as shown in Eq. (6), while a four-body feature can be obtained by $\boldsymbol{\psi}_{i,l_1}\boldsymbol{\psi}_{i,l_2}\boldsymbol{\psi}_{i,l_3}$. The factorial factor in Eq. (6) is just the degeneracy factor of the corresponding element in the tensor. Indeed, the Gaussian Moments Neural Network (GM-NN) proposed by Zaverkin and Kastner constructs the GM descriptor of many-body interactions by tensor contraction of GTOs, including part of (incomplete) four-body and higher order terms.[147]

Similarly, Drautz developed the Atomic Cluster Expansion (ACE) method[148] in the spherical harmonics representation more clearly, which efficiently calculates many-body interactions of an arbitrary order through tensor contraction. Ceriotti *et al.* proposed the *N*-body Iterative Contraction of Equivariants (NICE) method[149], which iteratively calculates high-order many-body interactions. Csányi *et al.* proposed the atomic PIP (aPIP) method[150], which extends the concept of PIPs for the entire system to the local environment. The aPIP method avoids the combinatorial explosion in the number of PIPs by retaining many-body terms within the cutoff radius, and can be considered an equivalent representation of MTP[146] and ACE[148]. In addition, an approach similar to ACE was taken by Fan *et al.* to generate angular descriptors with three-body to five-body terms combined with NNs in developing their neuroevolution potential (NEP) model[151], manifesting a high accuracy and super low cost. A more detailed derivation and the connections among these methods can be found in the comprehensive review of Ceriotti and coworkers[51]. It is worth noting that although the computation of an individual term scales linearly with $N_c$ for a given body-order in these advanced methods, the combinatorial growth of the necessary terms with increasing body-order makes the calculation of high order interactions still intractable. Consequently, one has to truncate the interactions to lower orders, *e.g.* four-body or five-body, in practical applications[148-150]. To date, how to efficiently construct an optimal complete set of local many-body features for a general system remains an open challenge[119]. Despite that, approaches based on conventional local descriptors have been most frequently used in constructing MLPs of extended systems. One limitation of such approaches is the lacking of description of non-local and long-range interactions, as discussed below. Another limitation is that the descriptor is expanded in basis functions in a fixed form, although their hyperparameters can be optimized by the automatic differentiation technique.

**C. Message Passing Neural Network-based Models**

More recently, message passing neural network[89] (MPNN)-based approaches have gained the increasing popularity for developing MLPs in an end-to-end manner. MPNN is a specific variant of the graph neural network (GNN)[152] model in the context of computer science[153]. In general, molecule is viewed as a graph $\{\mathbf{n}_i, \mathbf{e}_{ij}\}$, as shown in Figure 2(b), where each node $\mathbf{n}_i$ corresponds atom *i* and is connected to the neighbor atom *j* within a cutoff radius through the edge $\mathbf{e}_{ij}$, for which the interatomic distance $\mathbf{r}_{ij}$ is often used. A so-called message-passing (MP) process aggregates neighborhood information within a cutoff into the center layer-by-layer. A normal passing formalism for $\mathbf{m}_i^{t+1}$, the message of atom *i* in (*t*+1)th iteration, can then be concisely described as,

$$\mathbf{m}_i^{t+1} = \bigoplus_{j \in L_i} M^t\left(\mathbf{h}_i^t, \mathbf{h}_j^t, \mathbf{e}_{ij}\right) \tag{13}$$

$$\mathbf{h}_i^{t+1} = U^t\left(\mathbf{h}_i^t, \mathbf{m}_i^{t+1}\right) \tag{14}$$

where $\mathbf{h}_i^t$ is the hidden state feature of $\mathbf{n}_i$ at the $t$th iteration that captures its local information, messages are then passed between nodes along edges. $M^t$ is a learnable message function acting on the features of the nodes connected by the edge $\mathbf{e}_{ij}$. The $\bigoplus_{j \in L_i}$ is a differentiable and permutation invariant pooling operation over the set of neighbors $L_i$, such as sum, mean, or max to aggregate the message at each node $i$ to produce an updated message $\mathbf{m}_i^{t+1}$, which in turn is used to update the hidden feature $\mathbf{h}_i^{t+1}$ with the node update function ($U^t$) for the next MP cycle. In the readout phase, the atomic states are mapping into atomic energies with a learnable function $R^t$. It should be noted that most MPNN models use only one single readout function that takes the state after the last $T$th iteration and then maps it to the energy, namely $E_i = R^T(\mathbf{h}_i^T)$, while some models[154] use the atomic states from every iteration, as $E_i = \sum_t^T R^t(\mathbf{h}_i^t)$.

As illustrated in Figure 2(c), MPNN models refine the molecular representation by iteratively exchanging local environmental information between the central atom and its immediate neighbors, then between the neighbors and their neighbors, and so on. After multiple iterations of MP, the resulting structural features capture not only higher-order correlations within the original local environment but also interactions between atoms inside and outside the initial cutoff radius, thereby enriching the overall structural representation. Throughout this process, the dependence on any pre-defined descriptors is minimized, and all NN parameters and hyperparameters of basis functions in both the MP layers and the output layer are simultaneously optimized in an end-to-end manner. It is worth noting that as the number of MPs increases, the overall NNs become much deeper. Consequently, the use of residual blocks[155] in MPNNs has become very common, replacing the fully connected NN layers typically used in local descriptor-based models, which can alleviate issues such as vanishing or exploding gradients.

Early MPNN models[88, 154, 156, 157] were designed to pass two-body messages, in which the edges of the molecular graph encode only information about interatomic distances. The first MPNN-based MLP is the Deep Tensor Neural Network (DTNN) model[88] proposed by Schütt *et al.*, where messages are constructed based on a global distances matrix expanded in a Gaussian basis. Shortly thereafter, Schütt *et al.* introduced the SchNet[156] model as a variant of DTNN[88], replacing the Gaussian expansion of interatomic distances with atomic positions in the messages. SchNet[156] truncates the interatomic distances using a cutoff radius, making it applicable to systems with a large number of atoms, including periodic systems. Combining the MPNN concept with many-body expansion, Lubbers *et al.* proposed the HIP-NN model[154]. While message functions are also constructed by pairwise distances within a cutoff radius, HIP-NN[154] further decomposes the atomic energy into contributions from different many-body orders, which depend on the atomic features in each iteration of MP and can be used to estimate the uncertainty of model prediction. Unke *et al.* proposed PhysNet[157], in which interatomic interactions are modeled by learnable distance-based attention masks biased toward an exponential decay. In addition to learning atomic energy, PhysNet[157] has learned the partial atomic charge and the dispersion correction energy to include the electrostatic and dispersion interactions in the total energy expression.

MPNN models passing two-body features only have known to be incomplete to represent atomic structures.[158] Subsequent MPNN models, such as DimeNet[159] and Cormorant[160], started to incorporate angular information into the message functions, enhancing the representability while scaling quadratically with the number of neighboring atoms. Interestingly, SpookyNet[161] encodes angular information implicitly using basis functions expanded in Bernstein polynomials and spherical harmonics, achieving a linear scaling with respect to neighboring atoms. On the other hand, conventional local many-body descriptor-based models can be updated to an MPNN form to increase their flexibility. For example, Isayev *et al.* upgraded the ANI-1[132] model to the MPNN-based AIMNet[162] model by updating the two-body and three-body ACSFs associated weights through MP, thereby improving accuracy and transferability. We realized that the orbital coefficients in the EAD descriptor can be simply modified in a similar way and proposed a recursively EANN model, REANN for short[158]. Specifically, the orbital coefficient of the $j$th neighboring atom in the Eq. (6) can be rewritten as the output of an atomic NN ($U_j^t$) associated with that atom, which in turn depends on the EAD descriptor ($\boldsymbol{\rho}_j^t$) taking atom $j$ as the central atom,

$$c_j^{t+1} = U_j^t(\boldsymbol{\rho}_j^t(\mathbf{c}^t, \mathbf{r})) \tag{15}$$

where $\boldsymbol{\rho}_j^t$ are the EAD in the $t$th iteration and calculated with the relative coordinate vector $\mathbf{r}$ and corresponding coefficients $\mathbf{c}^t$ of all neighboring atoms centered on the atom $j$. By analogy, any neighboring atom of the atom $j$ can be taken as the center, and its orbital coefficient can be expressed similarly as Eq. (15), thus forming an iterative

MPNN architecture. This process is repeated until the last iteration, where the atomic NN will output the atomic energy.

These three-body feature-based MPNNs, such as REANN[158] and SpookyNet[161], significantly outperformed these models based on original three-body local descriptors (or even four-body ones), and two-body feature-based SchNet[156] on a representative $CH_4$ dataset[117] used to assess the completeness of atomic descriptors[163]. Additionally, because atoms outside the cutoff radius can implicitly correlate with those inside through MP, it is possible to use a smaller initial cutoff radius for the MPNN model and increase the layer of MPs to increase the effective cutoff, reaching the same fitting error compared to conventional local descriptor-based models like BPNN[82], as shown for a bulk water dataset[158]. However, Ceriotti *et al.* found that increasing the MP layer is less effective than directly increasing the cutoff radius in describing long-range interactions[164]. Jiang *et al.* found that a cutoff radius exceeding the first neighbor shell is necessary[165]. While increasing the MP layer can lower the average test errors of energy and forces in general, it may not improve predictions for structure/dynamical properties if the direct interatomic correlations within the local environment are missing[165].

A common feature of these MPNN models is that the atomic features being passed in every MP iteration are scalar quantities and invariant with respect to symmetry transformations. More recently, a new class of MPNN models passing equivariant features have been proposed, establishing the new state-of-the-art in the field. The idea is to exploit symmetry operations in feature construction rather than just making the features satisfying the symmetry. Instead of operating on these predefined invariant descriptors or features, equivariant MPNN models explicitly incorporate the relative orientations of interatomic pairs in the local environment, enabling superior representability of atomic structures and data efficiency, compared to the invariant counterparts, within the same cutoff radius.[166] An additional feature of the equivariant MPNN model is that it naturally represents the rotational equivariance of molecular response properties, such as dipole moment and polarizability.[43]

In past a few years, various schemes[159, 160, 166-176] have been proposed to generate equivariant features with different rotational orders (typically denoted by $l_{max}$, *i.e.*, $l_{max}$=0 for scalar, $l_{max}$=1 for vector, and $l_{max}\geqslant 2$ for tensor). A common way to construct equivariant features is based on the irreducible representation theory, achieving rotational equivariance subject to a specific group, such as SO(3), E(3) and others. As shown in Figure 3, a key step is that atomic features and messages are built on spherical harmonic tensors and coupled in an equivariant manner. Practically, this is often done through spherical tensor products via contraction with Clebsch-Gordan (CG) coefficients ($C_{l_1 m_1, l_2 m_2}^{l_3 m_3}$) to generate features with the desired symmetry, for example,

$$\mathbf{m}_{i,l_3 m_3}^{(t+1)} = \sum_{l_1 m_1, l_2 m_2} C_{l_1 m_1, l_2 m_2}^{l_3 m_3} \sum_{j \neq i}^{N_c} \mathbf{R}_{l_1}^{(t+1)}(\mathbf{r}_{ij}) \mathbf{Y}_{l_1 m_1}(\hat{\mathbf{r}}_{ij}) \otimes \mathbf{h}_{j,l_2 m_2}^{(t)} \qquad (16)$$

Pioneering models such as TensorField[177] and *N*-Body Networks[178], presented the proof-of-concept on how to construct equivariant features for MP within the SO(3) group for a molecular geometry. However, neither models were applied in practice to the PES training. Cormorant[160] was one of the first equivariant MPNN models applied to representing PESs, but it only passes rank-one tensor messages. GemNet[167] extends

DimeNet[159] by incorporating dihedral angle information and passing first-order tensorial message. These models demonstrated the benefit of incorporating rotational equivariance in predicting properties of small molecule datasets such as QM9[179] and MD17[22]. Later, respecting E(3)-equivariance, Kozinsky and coworkers proposed a series of models, including NequIP[166], Allegro[168], and BOTNet[169], which demonstrate state-of-the-art accuracy across a wide range of molecules and materials. NequIP[166] is an equivariant MPNN model based on two-body feature tensors (*i.e.* $l_{max} \geqslant 1$), demonstrating excellent predictive accuracy and generalization. As shown in Figure 4(a), the equivariant NequIP models ($l_{max} \geqslant 1$) significantly outperform the invariant ones ($l_{max}=0$) in force predictions for the bulk water system, manifesting about two orders of magnitude higher data efficiency enabled by equivariant tensor features. Allegro[168] is a strictly local equivariant many-body MLP, in which iterative tensor products are performed only between the equivariant features of the centered atom and its neighboring atoms within the cutoff radius. This architecture enables efficient parallelization for MD simulations as purely local descriptor-based models, scaling up to 100 million Ag atoms[168]. As a refined body-ordered adaptation of NequIP[166], BOTNet[169] increases the body order with each iteration of MP while using a non-linear update function. Other E(3)-equivariant models have also been proposed, such as SEGNN[170], which passes two-body tensorial message. ViSNet[171] possesses the SE(3) equivariance including the dihedral information in message function, achieving a better accuracy than lower-order MPNNs. It directly calculates the geometric information from a direction unit which only sums the vectors from the target node to its neighbors once, thus featuring a linear scaling with $N_c$. Based on a Transformer architecture with self-attention mechanism, models with E(3)/SE(3)-equivariance such as Equiformer[172] and EquiformerV2[173] have been proposed to further enhance the accuracy and generalizability of MLP when training with big data. Csányi and coworkers proposed a O(3)-equivariant MACE model[174], which combines the ACE method[148] for generating high-order many-body interaction terms (truncated to four-body) with equivariant MP. As displayed in Figure 4(b), by using high-order many-body ($\nu$=3) messages, MACE[174] requires only twice MPs to achieve a high-level accuracy in force prediction for aspirin, much better than using only two-body messages ($\nu$=1). It is also shown that the equivariant adaption enables a more obvious promotion of model accuracy for ($\nu$=1) than for ($\nu$=3) features. It is worth noting that Smidt *et al.* developed an e3nn library[180] for the efficient implementation of equivariant operations, which largely simplifies the construction of equivariant MPNN models. This library has been used to construct models such as NequIP[166], Allegro[168], MACE[174], DetaNet[44], and etc.

Although equivariant MPNNs based on spherical tensors have achieved great success, the high computational cost of spherical harmonics and CG tensor products limits their efficiency. To address this challenge, equivariant MPNN models with message tensors constructed in the Cartesian space have been proposed[43, 181-186]. In this regard, PaiNN[43], NewtonNet[181] and torchMD-Net[182] directly utilized vectorial messages in the Cartesian space (corresponding to $l_{max}$=1 in the spherical harmonic space). These models do not require couplings through CG coefficients. The tensor embedded atom network (TeaNet)[183] extended the MP procedure to incorporate rank 2 tensors in a very deep (16 layers) network. The Cartesian atomic cluster expansion (CACE)[184] model rederive the ACE framework in Cartesian space, to get polynormially independent invariant features in Cartesian coordinates directly. In the High-order Tensor Passing Potential (HotPP)[185] method, arbitrary *n*-order Cartesian tensors are used as messages with E($n$)-equivariance, enabling the prediction of corresponding tensor properties, such as dipole moments and polarizability. A recent Cartesian Atomic Moment Potential (CAMP)[186] model was also developed entirely in the Cartesian space, employing atomic moment tensor in combination with tensor products to incorporate higher body-order interactions.

Although tensor products can explicitly incorporate many-body interactions, the aggregations of interatomic correlations are generally along a single direction of MP. Recently, Wu *et al.*[187] developed an effective equivariant REANN model, EquiREANN, which can more quickly propagate the effective cutoff radius the by the MP procedure than commonly-used equivariant MPNNs. Messages in EquiREANN are constructed based on the superposition of equivariant tensors $\boldsymbol{\psi}_{i,\boldsymbol{l}}^{(t)}$, followed by a square operation that enables the expansion of the atomic environment in both directions simultaneously,

$$\boldsymbol{\psi}_{i,\boldsymbol{l}}^{(t+1)} = \sum_{j \neq i}^{N_c} \left( c_j^{(t+1)} \boldsymbol{\varphi}_{\boldsymbol{l}}(\mathbf{r}_{ij}) + \boldsymbol{\psi}_{j,\boldsymbol{l}}^{(t)} \right) \tag{17}$$

$$\rho_i^{(t+1)} = \boldsymbol{\psi}_{i,\boldsymbol{l}}^{(t+1)} \boldsymbol{\psi}_{i,\boldsymbol{l}}^{(t+1)} \tag{18}$$

The rotationally invariant EAD descriptor $\rho_i^{(t+1)}$ can then be used as input of the next MP to evaluate the orbital coefficients. A typical system highlighting the advantage of EquiREANN is the conjugated π-system like cumulene[48], in which the subtle potential energy varies due to the rotation of the $CH_2$ groups at both ends of the carbon chain.

This subtle energy variation is difficult to be accurately captured by invariant MPNNs such as SchNet[156], REANN[158], and even sGDML[86] based on global descriptors, as illustrated in Figure 5. EquiREANN keeps the simplicity of the REANN framework with nearly no extra computational cost, as the number of parameters remains unchanged. A drawback is that this model is less expressive without tensor contractions in each MP step.

It is important to note that Ceriotti *et al.* developed a general atomic structure representation theory based on atom-centered density, mathematically integrating local many-body descriptors, as well as invariant and equivariant MPNN models[164], into this framework, providing strategies for improving existing models. It is worth noting that MPNNs by construction adjust parameters in each iteration based on the last iteration, which complicates the design of multi-processing parallelization algorithms. The Allegro model bypasses this problem with restricting MP inside a local environment[168]. To improve the efficiency of equivariant MPNN models, Frank *et al.* proposed a self-attention based Euclidean equivariant transformer architecture called SO3KRATES, which separates invariant and equivariant information and replaces SO(3) convolutions with a filter on the relative orientation of atomic neighborhoods, thus eliminating the need for expensive tensor products[188].

**D. Nonlocality and Long-range Effects**

The atomistic representation of MLP is advantageous that its complexity is cast into atomic descriptors and truncated by the cutoff, making it easily extendable to high-dimensional problems. A drawback of this truncation is, however, that the non-local and long-range interactions beyond the cut-off are generally neglected. While this approximation is often reasonable, it is questionable for systems with extended charge transfer and strong van der Waals interactions, such as liquid-vapor interfaces[189]. While long-range interactions can be, in principle, achieved by increasing the cutoff radius, this treatment results in higher complexity of the atomic environment and more substantial computational cost, as the number of neighbor atoms increases cubically. Although increasing the number of MP layers in MPNNs can elongate the effective cutoff, there is always information lost in each MP step[164], making most MPNNs inefficient to deal with long-distance interactions, as shown in the cumulene case.[187, 190]

A more efficient way to describe long-range is to separate the short-range and long-range interactions from the total energy. This range separation scheme in MLPs was pioneered by Behler and coworkers[191, 192], as shown in Figure 6(a), where the total energy is given by,

$$E_{\text{total}} = E_{\text{short}} + E_{\text{elec}} = \sum_{i=1}^{N_{\text{atoms}}} E_i\left(\left\{\mathbf{G}_i\right\}\right) + \sum_{i>j}^{N_{\text{atoms}}} \frac{q_i\left(\left\{\mathbf{G}_i\right\}\right) q_j\left(\left\{\mathbf{G}_j\right\}\right)}{R_{ij}} \tag{19}$$

In practice, the short-range part is expressed by one BPNN[82], while another BPNN is introduced to predict atomic partial charges which also depend on the local environment, followed by an Ewald sum[193] of the Coulomb interactions between point charges[192]. Importantly, to avoid double counting of electrostatic energy contributions in both short- and long-range parts, electrostatic energies calculated by the first BPNN-based atomic charges are then removed from the total energies, leaving the remaining short-range part being trained with the other BPNN[82]. This strategy was also applied to augment other short-ranged MLPs like GAP[84] and SNAP[194], where long-range electrostatics are based on fixed atomic charges. Note that the reference partial charges (often DFT charges) used as targets for NN training in these models are not physical observables, which may depend on the chosen charge partitioning method. A slightly different treatment was used by Parkhill and coworkers in their TensorMol model[135], in which atomics charges are trained by NNs to reproduce dipole moments. Long-range physics was also included in a similar way in some MPNNs like PhysNet[157], in which atomic energies, partial charges, and dipole moments derived by atomic charges are all optimized by the same MPNN. Li and coworkers[195] developed a single charge-optimized BPNN model to simultaneously learn local atomic charges and atomic energies to match the total energy without relying on reference partial charges.

One limitation of this strategy is that it is unable to capture the global charge redistribution as atomic charges are typically given as functions of the local descriptors. Charge equilibration schemes have been often used to solve this problem in empirical force fields[196], which allow the global charge redistribution over the whole system to minimize the electrostatic energy with a constraint that the sum of partial atomic charges needs to be equal to the net charge of the system. This inspired Ghasemi *et al* to propose a charge-equilibration neural network (CENT) model.[197] In the CENT architecture, atomic partial charges are determined by the charge equilibration procedure[196] with atomic electronegativities being expressed by individual NNs whose inputs are local atomic descriptors. While CENT is suitable for describing ionic interactions, its total energy expression lacks of description of covalent bonding. An improved CENT model, referred to as CENT2[198], makes the short-range part independent of atomic charges and post-processes the charge density to obtain the trial electrostatic energy. More general MLPs combining CENT[197] and the standard local ML representation of short-ranged energy were also developed, for example, the fourth-generation high-dimensional neural network (4G-HDNNP)[199] and charge recursive neural network[200]. Both models require two atomic NN models, one for the atomic electronegativities used in charge equilibration yielding the globally optimized atomic charges and the other for atomic energies that sum up to short-ranged energy. Another related method is the Becke population neural network (BpopNN)[201], which uses modified SOAP descriptors encoding the atomic charge populations as the input of atomic NNs. Staacke *et al.* proposed a kernel charge equilibrium model, kQeq[202], which predicts environment-dependent electronegativity by kernel ridge regression and uses dipole moments as targets to avoid the ambiguity of charge partitioning schemes. Gironcoli and coworkers[203] further used a second-order Taylor expansion of the short-range energy and replaced two NNs in the 4G-HDNNP[199] framework with a single NN predicting all environment-dependent expansion coefficients. These MLPs can be constructed to describe several charge states of a system via one model.

Alternative approaches going beyond the point charge based long-range physics have been also proposed, for example, by learning the maximally localized Wannier centers (MLWCs) instead of atomic charges[204, 205], potentially allowing a more flexible representation for electrostatics. Interestingly, the self-consistent field neural network (SCFNN) model[205] of Gao and Remsing incorporates electrostatics via two modules, each consisting of a short-range network and a long-range network. The first module uses MLWCs to predict electronic response upon an effective electric field. A self-consistent procedure is performed to converge MLWCs and the effective electric field. After this, the second module uses the field alongside the local structural descriptors to predict nuclear forces in which long-range polarization is considered. Another interesting method is the long-distance equivariant representation[206, 207], which is designed to describe electrostatic interactions between not only charges but also static dipoles. It essentially evaluates a "potential field" by convoluting the neighbor density with a suitable kernel encoding the Coulomb and other slow-decaying potentials, and then expanding it on an atom-centered basis to obtain local features for describing the long-range interaction[207].

In addition to electrostatics and dispersions, delocalized electrons in conjugated $\pi$-systems like aromatic compounds or long-chain cumulenes can also give rise to rather long-distance interactions. Such cases are challenging even for global descriptor-based MLPs, *e.g.* sGDML[86], where the variation of the distance between these terminal atoms with their relative orientation is likely too weak to be captured by the global descriptor. A self-attention based MPNN model, SO3KRATES[188], was designed to solve this problem by transforming the MP iteration from the Cartesian coordinate space to the spherical harmonic coordinate space, where long-distance interactions in the Cartesian coordinate space can be very close in the spherical harmonic coordinate space.[190] Liu and coworkers recently developed the many-body function corrected neural network approach (MBNN)[208], which expands the total energy in a MB form and could be coupled with atomic attention (MBNN-att) [209], as illustrated in Figure 6(b). Importantly, instead of directly outputting atomic energies, these atom-centered NN outputs correspond to parameters of empirical MB functions, which are assigned with a much longer cutoff (*e.g.* larger than 10 Å) than typically used for the atomic descriptor (*e.g.* 6 Å), allowing for a more efficient description of long-range interactions. To capture the complex long-range intermolecular interactions, Hu *et al.* developed a MS-MACE model[175] using invariant MACE ($l_{max}$=0) in long-range and equivariant MACE ($l_{max}$=2) in short-range separately, reducing the overall cost while keeping the high accuracy of the short-range potential.

Finally, it should be noted that even for global descriptor-based MLPs, it is also useful to incorporate a physical expression of the long-range interaction to extrapolate the potential to the asymptotic region where ab initio data are absent. In this regard, a physics-based ML model, MLRNet[210], has been developed, where the outputs of NNs are integrated into a physically realistic Morse/long-range (MLR) function making it extrapolatable in both short and long range two-body interactions.

## IV. Selected Applications and Universal Potentials

Thanks to the availability of many open source and user-friendly packages, MLPs have been increasingly applied in chemistry, physics, and materials science, making it possible to perform atomistic simulations of complex systems that were unimaginable not long ago. Table I summarizes numerous MLP packages and their available links. There have been quite a few reviews summarizing the applications of MLPs in different research fields[24, 31, 48, 51, 211], so only some representative results based on different types of MLPs in recent a few years are discussed below.

Global descriptor-based approaches like PIP[76], PIP-NN[106-108], and FI-NN[114] have become most popular for developing highly accurate PESs based on high-level *ab initio* calculations to study spectroscopy and reaction dynamics of small molecules.[19] For example, Li, Guo and coworkers constructed the PIP-NN PESs at the CCSD(T) level for the F/Cl + $CH_4$ reactive system, on which quantum-state resolved dynamics simulations revealed the reaction mechanisms of stereodynamic control[212] and Feshbach resonances[213] in these polyatomic reactions. Yang *et al.* developed an ab initio PIP-NN PES with the spectroscopic accuracy for the HF-HF system and performed full-dimensional quantum scattering calculations, demonstrating that HF-HF inelastic collisions do not follow the well-established energy and angular momentum gap laws.[214] Fu, Zhang, and coworkers applied the FI-NN method to construct the global PES of the $F^-$ + $(CH_3)_3Cl$ reaction, which was so far the largest system of bimolecular reactive scattering being studied with multiple reaction channels, as shown in Figure 7(a). This 39-dimensional FI-NN PES allows the authors to reveal in detail the competitive mechanisms between base-induced elimination and bimolecular nucleophilic substitution ($S_N2$) reactions, uncovering the dynamical origin of the very low $S_N2$ reactivity.[116] Very recently, Bowman and colleagues developed a 128-dimensional PIP PES to calculate infrared spectrum of a linear alkane $C_{14}H_{30}$. Numerous gauche configurations of this highly flexible molecule was identified and the global minimum was found irrelevant to the experimental spectrum[103]. Müller and colleagues constructed an sGDML PES for a double-walled nanotube containing up to 370 atoms, which, to the best of our knowledge, is the largest molecule studied to date using global descriptor-based methods.[126] As illustrated in Figure 7(b), with nanosecond-long classical and path-integral MD simulations, their findings revealed the nuclear quantum effects and long-range effects on the rotation of nanotubes.

In molecule-surface systems, local descriptor-based methods like BPNN[82] and EANN[142] have become the mainstream schemes for constructing high-dimensional PESs involving surface atoms[215] accounting for energy transfer upon gas-surface collisions[216-218]. These MLPs can uniformly capture interactions between molecules and different surfaces, as well as between molecules and surfaces at varying coverages and sizes. For example, Gerrits constructed a BPNN[82] PES to describe the dissociative adsorption of $D_2$ molecules on a curved Pt crystal with multiple facets or a continuous change of step density, revealing the step-type correlated reaction mechanisms and a shadow effect of the stepped sites[219]. Alducin, Juaristi, and coworkers developed an EANN PES to describe the laser-induced desorption and oxidation on Pd(111)[220] and Ru(0001)[221] of multiple CO adsorbates, monitoring the dynamic changes of CO coverage during the reaction induced by hot electrons and thermal phonons. Gu *et al.* constructed an EANN PES for the H spillover process on a Pt/Cu(111) single atom alloy surface, demonstrating that collisions between $H_2$ and adsorbed H atoms at the Pt site play a crucial role in driving this process.[222] More recently, Stark *et al.* benchmarked the performance of several different types of MLPs in the $H_2$/Cu(111) system and found that REANN and MACE models provide the best balance between accuracy and time-to-solution.[223] While ACE models are fastest, they were not able to achieve sufficiently accurate predictions of gas-surface dynamics.[223]

In condensed phase systems, various atomistic MLPs have gradually replaced conventional empirical force fields and are now widely used to investigate structure, phase transitions, and thermodynamic properties. For example, Csányi and coworkers developed several general-purpose SOAP-GAPs for single elements such as C, Si, and P.[224, 225] These MLPs can describe a broad range of observable structures, including crystal, liquid, and amorphous phases, demonstrating both accuracy and transferability. They also enabled efficient first-principles MD simulations of dynamical processes, such as self-diffusion, amorphous formation, and phase transition in these single-element materials. Multi-element materials have been also studied by MLPs. For example, Yin *et al.* developed an MTP potential to study the mobility of edge and screw dislocations in a body-centered cubic MoNbTaW refractory high-entropy alloy (RHEA) across a wide temperature range.[226] They revealed the salient mechanisms and the specific effects of temperature and local chemical order on the motion of edge and screw dislocations. Bulk water has been an important condensed phase system to benchmark MLPs. In this respect, Behler, Ceriotti, Car and other research groups have successively constructed BPNN and DP models for bulk water at the DFT level[227-231]. These models have been used to simulate thermodynamic properties such as self-diffusion coefficient, density, self-dissociation constant, and phase diagram, revealing the underlying physical mechanisms behind the anomalous thermodynamic properties of water at the first-principles level. Using the DP potential, Lin *et al.* studied phase transitions of monolayer water/ice in nanoconfinement with hydrophobic walls. They identified two previously unreported high-density ice phases and a negative pressure region for the low-density monolayer ice phase, thus enriching the phase diagram of monolayer water/ices.[231]

One of the main advantages of MLPs is their ability to better describe bond formation and breaking, enabling the analysis of rare reaction events. This has led to their great success in simulating complex chemical reactions in the condensed-phase that were challenging for conventional force fields. For example, Zeng *et al.* developed a DP potential for methane combustion based on *ab initio* data, sampled from reactive force field trajectories *via* AL.[232] This MLP enabled a complete reaction network for methane combustion and identified many new reactions not included in experimental databases. Galib and Limmer constructed a DP potential to study the uptake of $N_2O_5$ into an aqueous aerosol. The elementary physical and chemical steps involved in the reaction are illustrated in Figure 7(c), accompanied with simulation snapshots. It was identified that the uptake of $N_2O_5$ is not mediated by the bulk but rather dominated by interfacial processes, offering a reasonable explanation for existing experimental observations.[233] By combining the ANI model[132] with active learning, Smith and colleagues developed ANI-1xnr, a general-purpose reactive MLP model for C, H, N, and O elements in the condensed phase.[234] This model facilitates high-throughput in silico experiment in reactive chemistry, with applications including carbon solid-phase nucleation and the Miller experiment.

MLPs have also been extensively applied to heterogenous catalytic reactions, significantly extending the time and spatial scales of catalysis-related researches. Due to the complex structural changes of catalysts and the variety of elementary reactions, these MLPs must cover a larger chemical configuration space. By combining the global optimization algorithm of stochastic surface walking (SSW)[235] with the local descriptor-based NNPs[130], Liu, Shang, and colleagues developed the LASP software[236] and constructed a large number of NNPs for many important catalysts and catalytic reaction networks[31]. For example, they established a thermodynamic phase diagram for ternary zinc-chromium oxide (ZnCrO), shedding light on the mechanism of catalytic syngas conversion over ZnCrO catalysts at different Zn:Cr ratios, as shown in Figure 7(d).[237] With the acceleration of NNPs, they also performed large-scale grand canonical structure exploration for the silver-catalyzed epoxidation of ethylene, identifying a unique Ag surface oxide phase, *i.e.* the $O_5$ phase, as the active phase for ethene epoxidation under industrial catalytic conditions.[238] In addition to the SSW-NN scheme, other MLPs and sampling schemes have been developed. For example, using a global optimization scheme based on BPNN potential, Behler and colleagues studied the copper clusters at the ZnO surface and their extension to the ternary CuZnO system, identifying a series of structures with common structural features resembling the Cu(111) and Cu(110) surfaces at the metal-oxide interface.[239] Xu *et al.* have integrated Grand Canonical Monte Carlo (GCMC) simulations with an EANN potential to study the oxidation of large-scale flat and stepped PtOx surfaces during the catalysis process in real conditions.[240] They not only identified several key PdOx intermediates among a huge number of local structures, consistent with experimental observations, but also revealed the mechanism of surface oxide formation on Pt without the need to manually construct a surface model. Yang *et al.* combined the PaiNN[43] potential and enhanced sampling to investigate the reactive process of the oxygen reduction reaction (ORR) at an Au(100)-water interface, identifying the associative reaction mechanism without the presence of *O and a low reaction barrier of 0.3 eV, which well explains the outstanding experimental ORR activity.[241] Gong *et al.* developed a DP potential[242] for the dissociation process of a $CO_2$ molecule on copper nanoclusters, unveiling an anomalous entropic effect on catalysis via surface pre-melting of nanoclusters.[243] Bunting *et al.* developed a NequIP model[166] for the dehydrogenation of propane on the surface of copper nanoparticles doped with single-atom Rh or Pd, and revealed a profound effect of the dynamics of these structures on the calculated catalytic activity of single-atom alloys.[244] Yang and Parrinello developed a DP model[242] that enables enhanced dynamics simulations of the decomposition process of $NH_3$ on $Li_2NH$ surface catalysts, explaining the high-temperature stability of industrial catalysts and unraveling the complex dynamic behavior of catalytic processes.[245] These results demonstrate the current capabilities of MLPs to investigate realistic in-situ catalyst structures and the reaction mechanisms of heterogeneous catalysis.

MLPs have also found significant applications in the simulation of energy materials. For example, various MLP methods, including GNN, MTP, DP, and EANN, have been applied to accelerate MD simulations of solid electrolytes at low temperatures in place of the expensive AIMD simulations.[246-249] Some of these MLPs have been able to uncover the change of diffusion mechanisms because of the ordered-disordered structural transformation in super ionic conductors. Using DP-based MD simulations, Lin *et al.* investigated the chemical shifts in paramagnetic battery materials, revealing fast alkali-ion dynamics and achieving excellent agreement with experimental measurements.[250] As shown in Figure 7(e), Ong and coworkers used the MTP to comprehensively study the thermodynamics and kinetics of the cathode-electrolyte interface in all-solid-state Lithium-sulfur batteries, demonstrating the formation of $Li_xS_y$ and $S_x$ species at the interface.[251] Additionally, various MLP methods, such as DP, ANI-1, GAP, and Allegro, have been applied to simulate ionic liquids, demonstrating accuracy in predicting the structural, thermophysical, and transport properties.[252-255]

MLPs have been recently started to be applied to biomolecular systems. Müller and colleagues developed SpookyNet[161] potentials for polyalanines and a 46-residue protein crambin in aqueous solution with 8205 explicit water molecules (>25,000 atoms), enabling nanosecond-scale MD simulations at essentially *ab initio* quality. [256] Their findings revealed previously unknown intermediates in the folding pathway of polyalanine peptides and a dynamical equilibrium between α- and $3_{10}$-helices, suggesting that simulations at *ab initio* accuracy may be necessary to understand dynamic biomolecular processes. Around the same time, Inizan *et al.* proposed a hybrid strategy that combines the ANI MLP model[132] for solute-solute interactions with a classical polarizable force field for solvent-solute and solvent-solvent interactions, enabling nanosecond MD simulations of biosystems containing approximately 100,000 atoms.[257] Based on the Allegro[168] model, Kozinsky and coworkers performed stable nanoseconds-long simulations of protein dynamics and scale up to a 44-million atom structure of a complete, all-atom, explicitly solvated HIV capsid.[258] Very recently, Wang *et al.* constructed the $AI^2BMD$ model, a VisNet-based MLP, to simulate full-atom large biomolecules, as illustrated in Figure 7(f), achieving generalizable *ab initio* accuracy for energy and force calculations of various proteins with more than 10,000 atoms.[259] Through several hundred nanoseconds of dynamics simulations, they were able to efficiently explore the conformational space of peptides and proteins, deriving accurate 3*J* couplings that match nuclear magnetic resonance experiments.

Most MLPs discussed above are customized for specific systems with a limited number of elements, making them difficult to transfer to other systems. Recently, this challenge has started to be addressed by the development of universal potentials (UPs), which, in principle, can be applied to any system, provided that they were well trained with very diversely distributed datasets. Table II summarizes numerous UP models and their training details, including the model architecture, model size, training dataset and number of chemical elements covered. These UP models have been greatly facilitated by the generation of big data in chemistry and materials science, for example, the Materials Project (MP) dataset[260] including diverse configurations with 89 elements largely across the periodic table and electronic structure properties. In this aspect, Ong *et al.* proposed the MatErials Graph Network (MEGNet)[261], probably the first UP model designed for material property predictions of inorganic crystals. MEGNet was first trained on ~60,000 minimum energy configurations with formation energies from an initial version of the MP dataset, which then demonstrated good transferability by transfer learning over ~10% of these structures with elastic constants. However, MEGNet was not trained with force and stress, thus lacking energy and force continuity and not suitable for MD simulations.

More recently, UP models capable of computing forces and stresses were reported based on augmented MP datasets, including M3GNet[262] and CHGNet[263], both of which were trained on snapshots along the trajectories of DFT relaxations of MP structures. M3GNet[262] was based on GNNs with three-body interactions, trained with energies, forces and stress over 187,000 configurations. With M3GNet, ~1.8 million materials were identified as potentially stable from a screening of ~31 million hypothetical crystal structures[262]. This result demonstrates the possibility for discovering synthesizable materials with desired properties using such UP models. By combining M3GNet and the DImensionality-Reduced Encoded Clusters with sTratified (DIRECT) sampling, this model was updated to M3GNet-DIRECT with improved extrapolability[264]. CHGNet[263] is a more expressive GNN model based on that explicitly incorporates magnetic moments for capturing the variability of chemical interactions across different valence states. It was pretrained on the Materials Project Trajectory (MPtrj) dataset[263] consisting of over 1.58 million inorganic structures across 89 elements. CHGNet[263] enables longtime ML simulations providing direct charge information and insights into ionic systems. Also based on the MPtrj dataset[263], Csányi *et al.* developed an equivariant UP model, MACE-MP-0[265], which utilizes the MACE[174] architecture and is claimed to be a foundation model for atomistic simulations. MACE-MP-0 is able to provide reasonable accuracy and show great transferability across diverse examples[265], including molecules, solid, liquid, and interfacial systems, even heterogeneous catalysis and combustion reactions, as shown in in Figure 8. Very recently, Cubuk *et al.* proposed the GNoME model[266] based on the equivariant NequIP[166] architecture. Impressively, an extended dataset was used in the training process, containing 89 million inorganic crystal structures, nearly two orders of magnitude bigger than the number of data of MPtrj[263]. This GNoME model enabled the efficient discovery of 2.2 million stable crystals with some out-of-distribution capabilities to find stable materials unknown to humanity[263].

Those UP models based on relaxation trajectory datasets are still limited by their generalizability across the vast configuration and chemical space, particularly under realistic conditions involving high temperatures and pressures. To address this challenge, Lu *et al.* developed the MatterSim model[267], which was trained with an extensive dataset covering a wide range of material structures sampled under varying temperatures (0 to 5000 K) and pressures (up to 1000 GPa). With a total of 182 million parameters, the largest model size to date, MatterSim achieved up to an order-of-magnitude increase in the prediction accuracy of energies, forces, and stresses for off-equilibrium material structures, compared with previous UP models.[267] Combining previous datasets like MP[260], MPtrj[263], and Alexandria[268], the Open Materials 2024 (OMat24)[269] dataset contains 118 million diverse structures and elemental compositions for inorganic bulk materials, appearing to be the largest dataset so far. A pre-trained model based on the OMat24[269] dataset using the EquiformerV2[173] architecture has achieved the state-of-the-art performance on the Matbench Discovery leaderboard[270].

Other UP models based on datasets beyond MP and its related databases have been also reported. For example, the ALIGNN-FF[271] model was trained on JARVIS-DFT[272], a DFT database of 75,000 inorganic crystals covering 89 elements, for data-driven materials design. Ibuka *et al.* proposed the PreFerred Potential (PFP) model[273], using the TeaNet[183] architecture and trained with a dataset of 9 million configurations. PFP was capable of modeling a diverse range of phenomena, including lithium diffusion and material discovery for Fischer-Tropsch catalysts[273], and was later expanded from 45 to all 96 long-lived elements[55]. Later, Wang *et al.* proposed DPA-1[274] using the DP[242] architecture with a gated attention mechanism. Pre-trained on the OC2M dataset[275] containing 56 elements, DPA-1 has been successfully applied to various systems, including high-entropy alloys. To increase the model generalizability, Wang *et al.* developed the DPA-2 model[276] by pre-training on a diverse set of chemical and materials systems with varying DFT settings, covering 73 elements. DPA-2 has shown a superior performance and generalizability than DPA-1[274] in a wide range of systems[276], including alloys, semiconductors, battery materials, and drug molecules. Additionally, a graph-based pre-trained transformer force field (GPTFF)[277] was trained on the Atomly database[277] containing approximately 37.8 million structures, which are more diverse than the MPtrj[263] dataset. GPTFF[277] demonstrates its ability to simulate arbitrary inorganic systems with high precision and generalizability. Very recently, Song *et al.* proposed the universal NEP (UNEP-v1) model[278] for 16 elemental metals and their diverse alloys, using the NEP[279] architecture and trained with a dataset of 105,464 configurations. The UNEP-v1 model[278] exhibits superior performance than the EAM force field across various physical properties and remarkable computational efficiency on GPUs. Beyond applications in materials, UP models have also been developed for molecules. Utilizing the ANI architecture, Smith *et al.* developed the ANI-1ccx model for organic molecules, demonstrating excellent performance for predicting thermochemical properties.[280] This model has been reformulated as ANI-1ccx-gelu by using the GELU activation function by Dral and coworkers[281], enabling the calculation of Raman and infrared anharmonic frequencies.

**V. Conclusion and Outlook**

MLPs have become one of the fastest-growing and most widespread applications of machine learning in chemistry. After years of development, various MLP methods have achieved a considerable level of maturity, as summarized in this review. However, open challenges and opportunities remain. For small molecules and reactions where the highest possible accuracy and efficiency are required, global descriptor-based models remain the mainstream MLP approaches. For example, the PIP PES for aspirin is found to be more accurate than that developed by ANI, GAP, and sGDML, and approximately 40 to 100 times faster than that by ACE when they achieve comparable accuracy.[95] In NN-based methods, it is crucial to develop more efficient algorithms to generate symmetry-adapted descriptors. A new algorithm proposed by Hao *et al.* very recently based on graph connectivity analysis has tremendously accelerated the generation of FIs over existing methods, which is very promising to extend FI-NN to more than 15-atom systems with high symmetry demands.[282] The resultant FI-NN PESs for several molecules in the MD17 dataset[22] achieve significantly smaller test errors than some local descriptor-based and MPNN models. While Kernel-based methods, such as pKREG[87] and sGDML[126], have the potential to be more data-efficient alternatives, their effectiveness in describing reactive systems across the global configuration space remains to be validated. Moreover, for such systems, it is possible and sometimes necessary to calculate diabatic (or quasi-diabatic) states for electronically nonadiabatic dynamics simulations. However, very limited ML-based methods have been developed to represent the diabatic potential energy matrix[283-286]. Accurately describing the global quasi-diabatic PESs for realistic polyatomic molecules with more than three electronic states remains a big challenge.[287]

Atomistic MLP methods have made significant successes in simulating extended systems over the past decade. However, in most cases, the many-body description of atomic structures is truncated to a finite order. Indeed, constructing an efficient and complete representation of complex atomic structures remains an open challenge.[119] While equivariant MPNNs have so far demonstrated high representability and generalizability, multiple tensor contraction operations make them computationally expensive. More efficient algorithms to enable low-cost equivariant MPNNs are highly desirable. In addition, the intrinsic correlation between sequential MP steps within these MPNN models presents a challenge for massive parallelization in large-scale MD simulations. This necessitates the design of a more sophisticated parallelization algorithm than the conventional atom-wise method. Furthermore, while various strategies to describe long-range interactions have been proposed for atomistic MLPs, there is still much room for improving their efficiency. Recent work on integrating Ewald summation based MPNNs showed some promise in this direction[288]. Interestingly, MLPs under external electric fields have emerged[289-291], allowing one to study the field-induced polarization and electrostatics.

A new trend of developing universal potentials across molecules and materials is just emerging and rapidly advancing. These universal models are often pre-trained on datasets focusing on crystals and materials, limiting their coverage of the vast chemical space. To make further progress, the training datasets should not only cover all elements in the periodic table but also the full combinatorial material space, including molecules, bulk solids and liquids, surfaces, interfaces, defects, and a broader range of non-equilibrium and reactive configurations. To this end, more efficient data sampling algorithms across chemical space are needed. Moreover, current architectures of universal potentials primarily rely on element-embedded MPNN or GNN models, typically with millions, or even tens to hundreds of millions, of parameters. In comparison to specialized models, they would exhibit much slower inference speeds and more memory demands. How to construct lighter universal potentials while keeping the accuracy level needs further exploration. In addition, while these universal potentials predict reasonably stable structures, their generalizability in atomistic simulations involving chemical reactions has yet to be fully validated. With continuous advancements in both big data and machine learning, we are optimistic that universal potential models will soon experience significant breakthroughs in the near future, broadening their applicability across a wide range of scenarios.

## Data availability

Data sharing is not applicable to this review, as no new data were created or analyzed in this study.

## Conflicts of interest

There are no conflicts of interest to declare.

## Acknowledgement

We thank the continuous support from the Strategic Priority Research Program of the Chinese Academy of Sciences (XDB0450101), Innovation Program for Quantum Science and Technology (2021ZD0303301), the National Natural Science Foundation of China (22325304, 22221003, and 22033007).

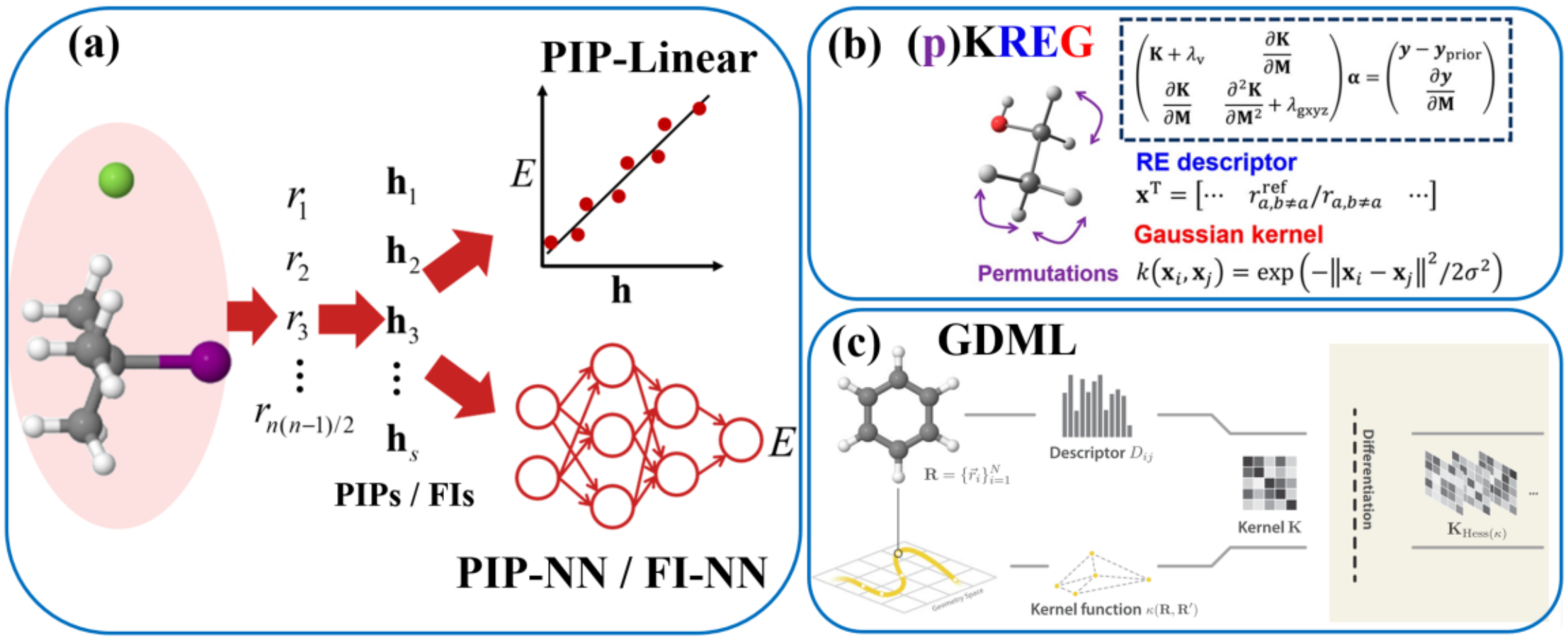

**Figure 1**. Schematic diagrams of several popular global descriptor-based MLPs, (a) linear regression (PIP-Linear) and NN (PIP-NN/FI-NN) models[76, 106, 114] using PIP/FI descriptors (**h**) generated from interatomic distances (***r***), (b) the (p)KREG model[125] using the ratio of relative-to-equilibrium (RE) distances as descriptor with proper permutational invariance of Gaussian kernel functions, reproduced with permission from Ref. 125. Copyright 2025 American Chemical Society, (c) the GDML model[22] using the Hessian matrix of a kernel that relates all interatomic distances as the covariance matrix in the kernel ridge regression, reproduced with permission from Ref. 22 under the Creative Commons CC-BY-NC license.

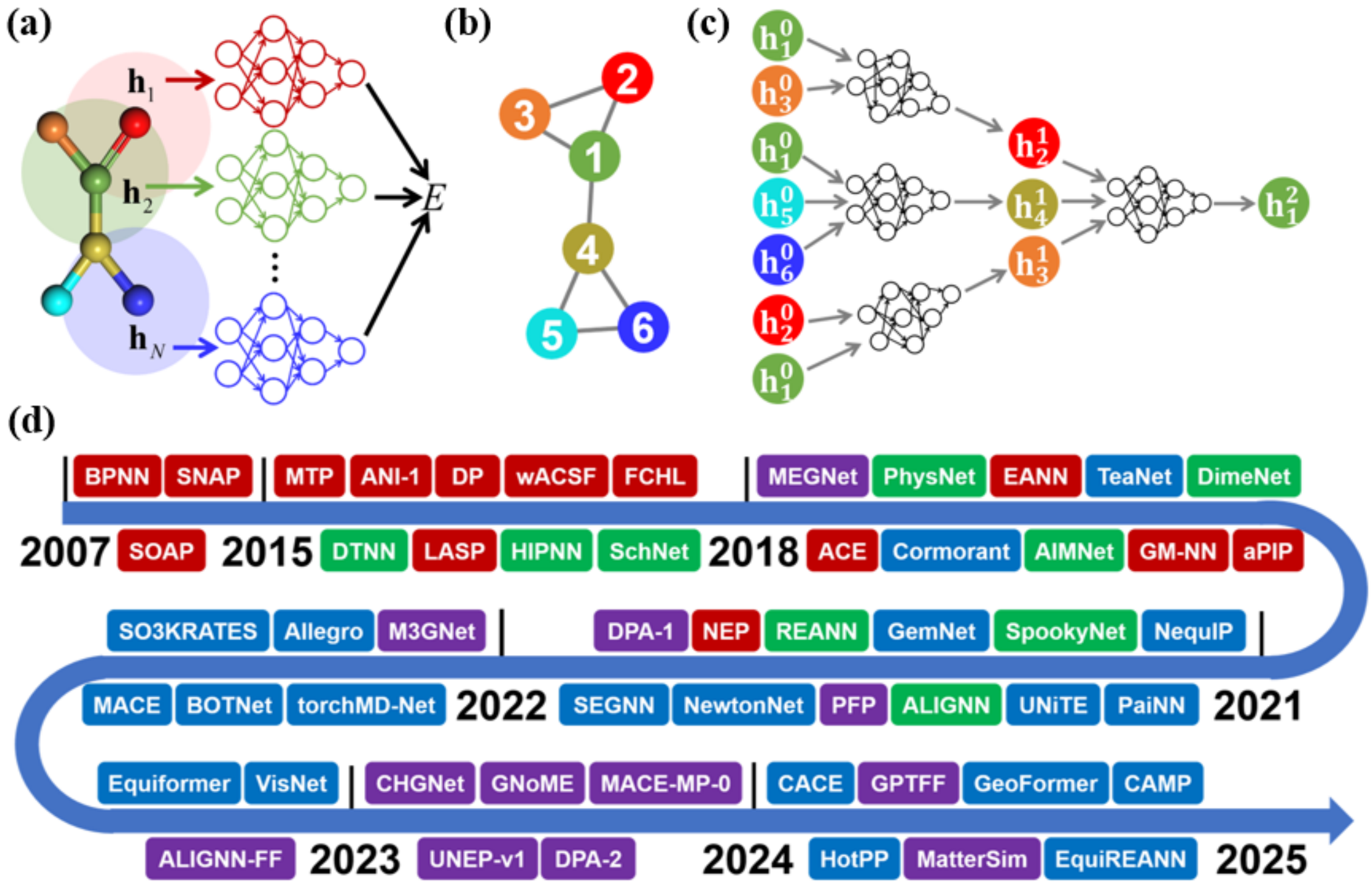

**Figure 2**. (a) A schematic of atomistic MLP models, (b) a molecular graph, where circles represent nodes corresponding to atoms and lines represent edges, (c) the generation of atomic features for the central atom 1 with *two* message-passing processes ($T$=2) and (d) the evolution of representative atomistic MLP models from 2007 to 2024, which are roughly sorted by their public release dates (*e.g.* an arxiv preprint if available). These MLP models are categorized as follows: strictly local descriptor-based models (red), invariant (green) and equivariant (blue) MPNN-based models, and universal potential models (purple). Recent a few years witness an increasingly fast development of atomistic MLPs.

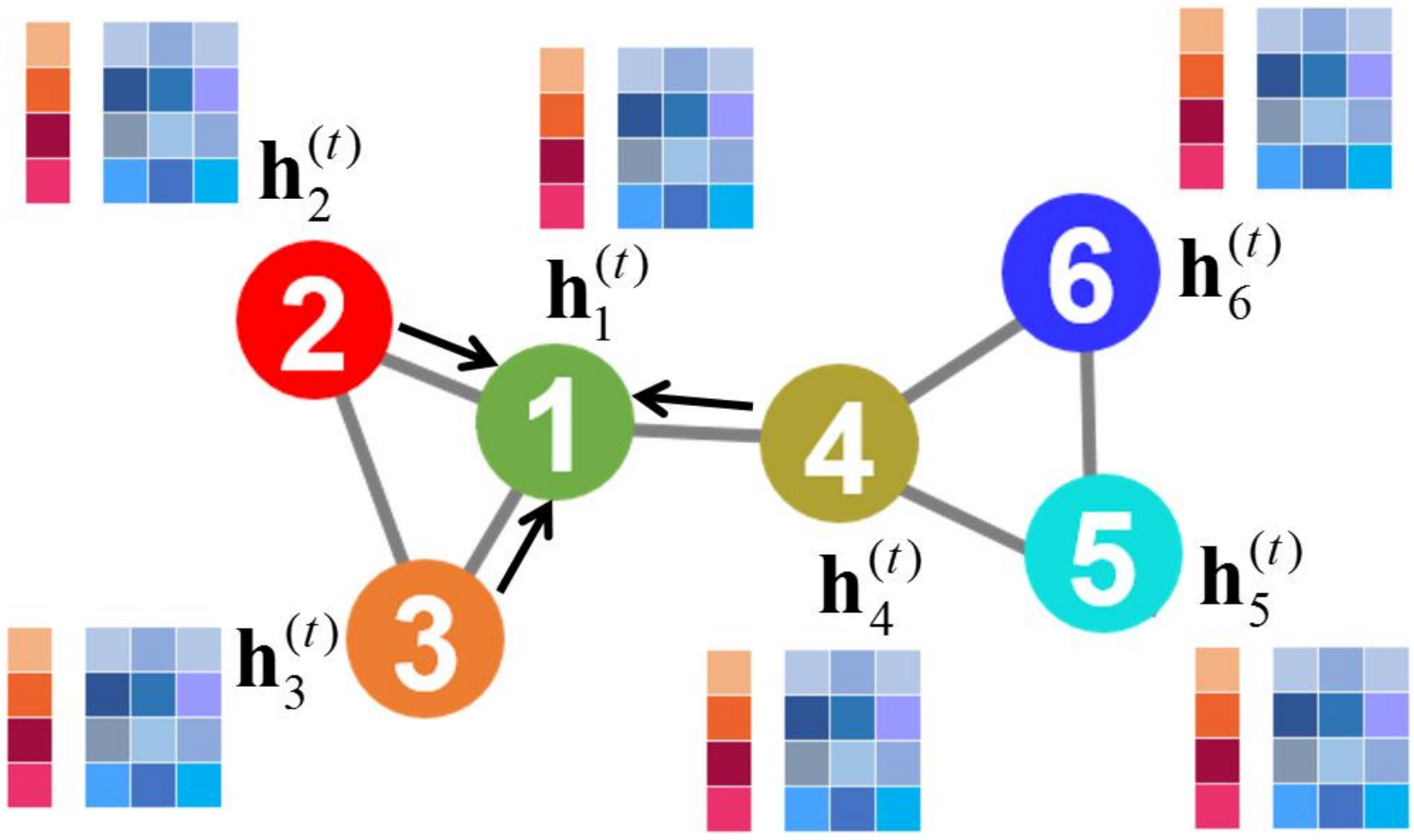

$\mathbf{h}_2^{(t)}$ $\mathbf{h}_1^{(t)}$ $\mathbf{h}_3^{(t)}$ $\mathbf{h}_4^{(t)}$ $\mathbf{h}_5^{(t)}$ $\mathbf{h}_6^{(t)}$

**Figure 3**. Schematic diagram for an equivariant MPNN, in which the message of atom 1 at the *t*th iteration is constructed based on the hidden state features of its neighboring atoms and each atom carries a feature vector composed of scalar features (4 red squares), vectorial features (4×3 blue squares), and possible higher-order tensorial features.

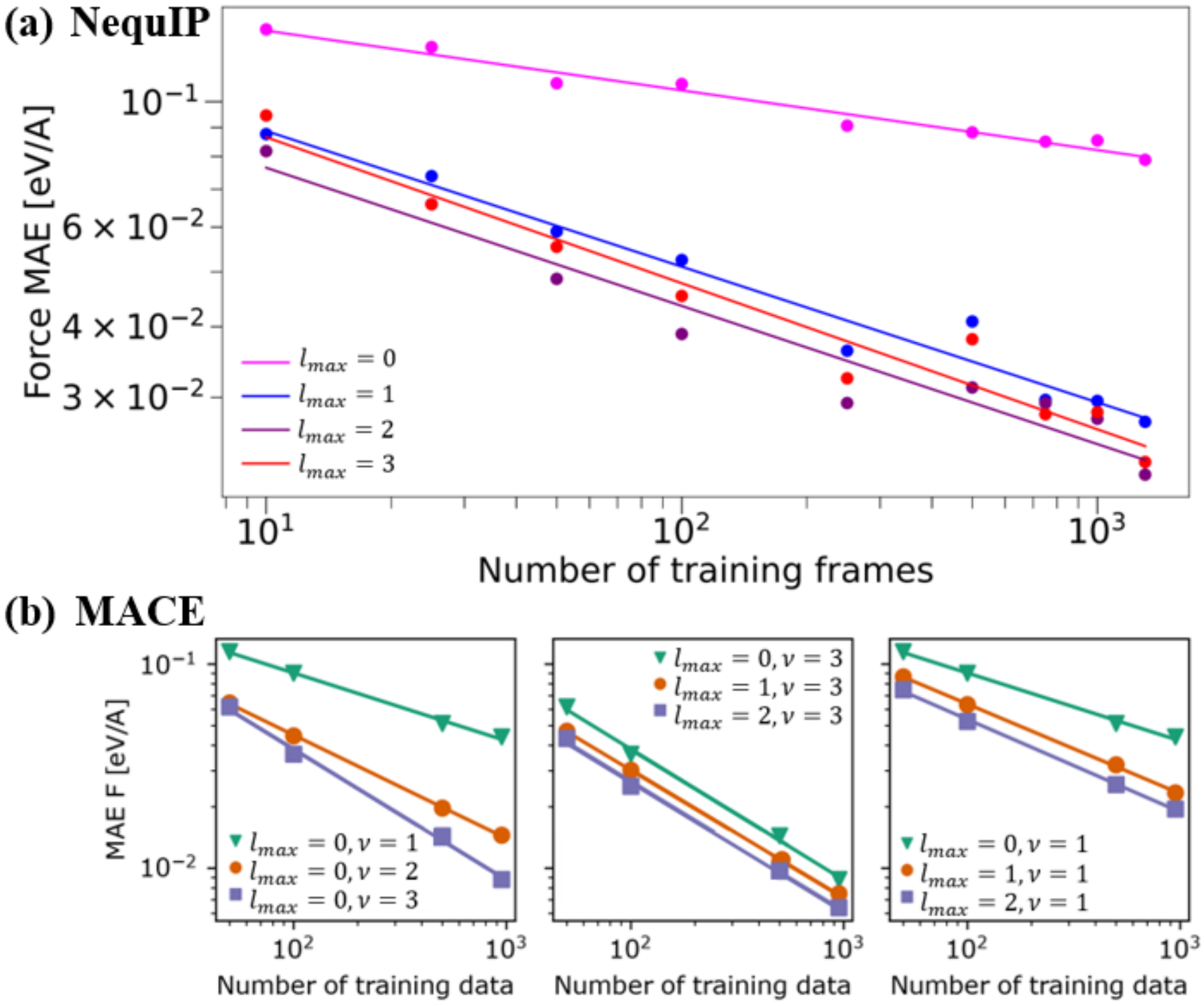

**Figure 4**. Learning curves of force errors of (a) NequIP[166] ($T$=6) and (b) MACE[174] ($T$=2) potentials that are trained for the bulk water and aspirin systems, respectively, shown with the varying rotation-order and/or body-order, as a function of the number of training data. (Adapted with permission from Ref. 166 under the CC BY 4.0 License and Ref. 174 with the author's authorization.)

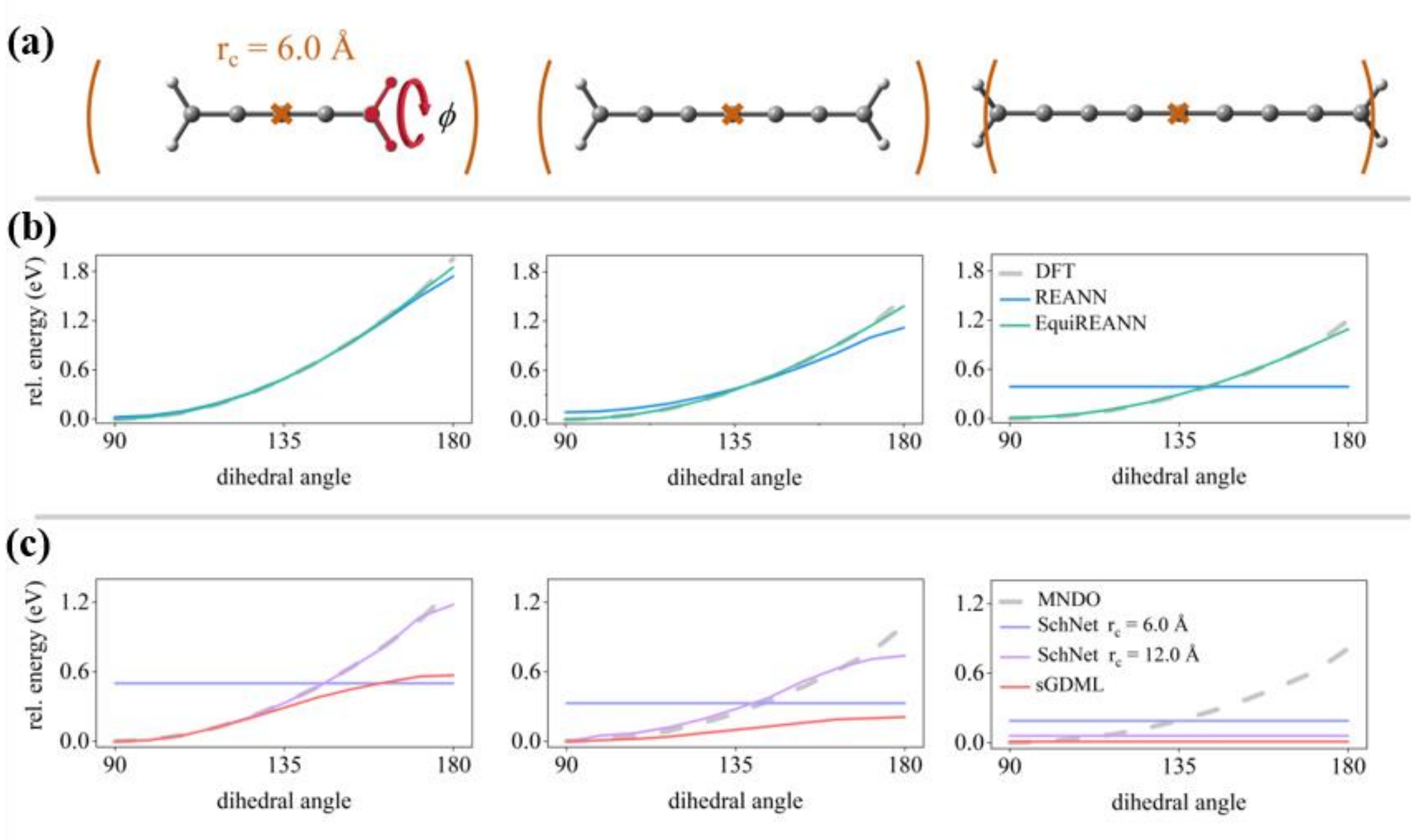

**Figure 5.** (a) Schematic diagrams of representative cumulenes, $C_5H_4$, $C_7H_4$, and $C_9H_4$, along with their respective cutoff spheres centered at the middle carbon atom. (b-c) Energy profiles as a function of the dihedral angle ($\phi$) in these cumulenes calculated with REANN[158], EquiREANN[187], SchNet[156] and sGDML[86]. The cutoff is set to 6.0 Å and illustrated by the orange curves. Note that the reference data have been generated with different methods (DFT or MNDO[292]) for (b) and (c), respectively, which however does not affect the comparison of the energy trend. Reproduced with permission from Ref. 187. Copyright 2025 American Chemical Society.

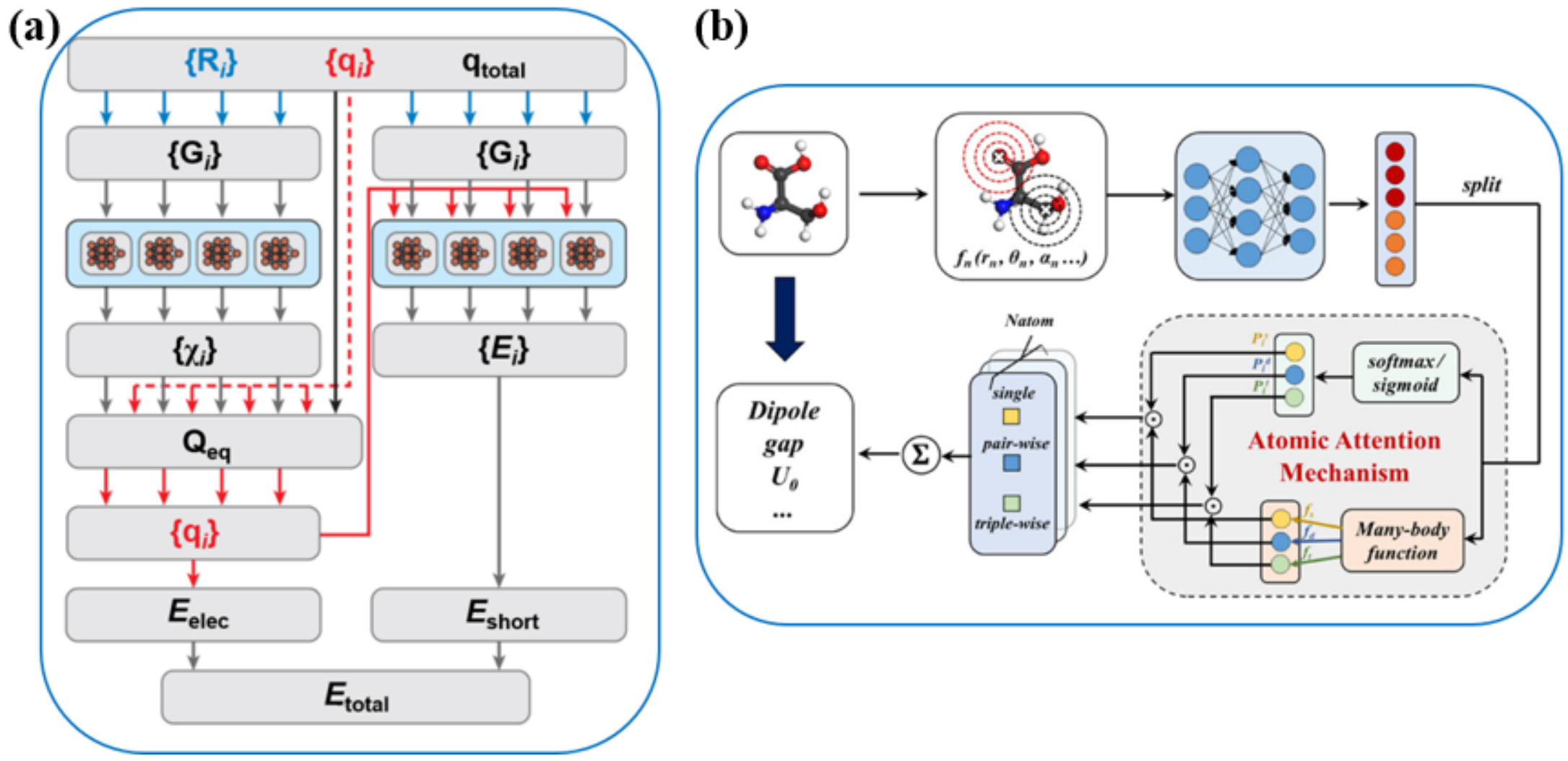

**Figure 6**. Two recent strategies to include long-range interactions in the atomistic MLP framework (a) the 4G-HDNNP model[199], reproduced with permission under the CC BY 4.0 license and (b) the MBNN-att model[209], reproduced with permission from Ref. 209, Copyright 2025 American Chemical Society.

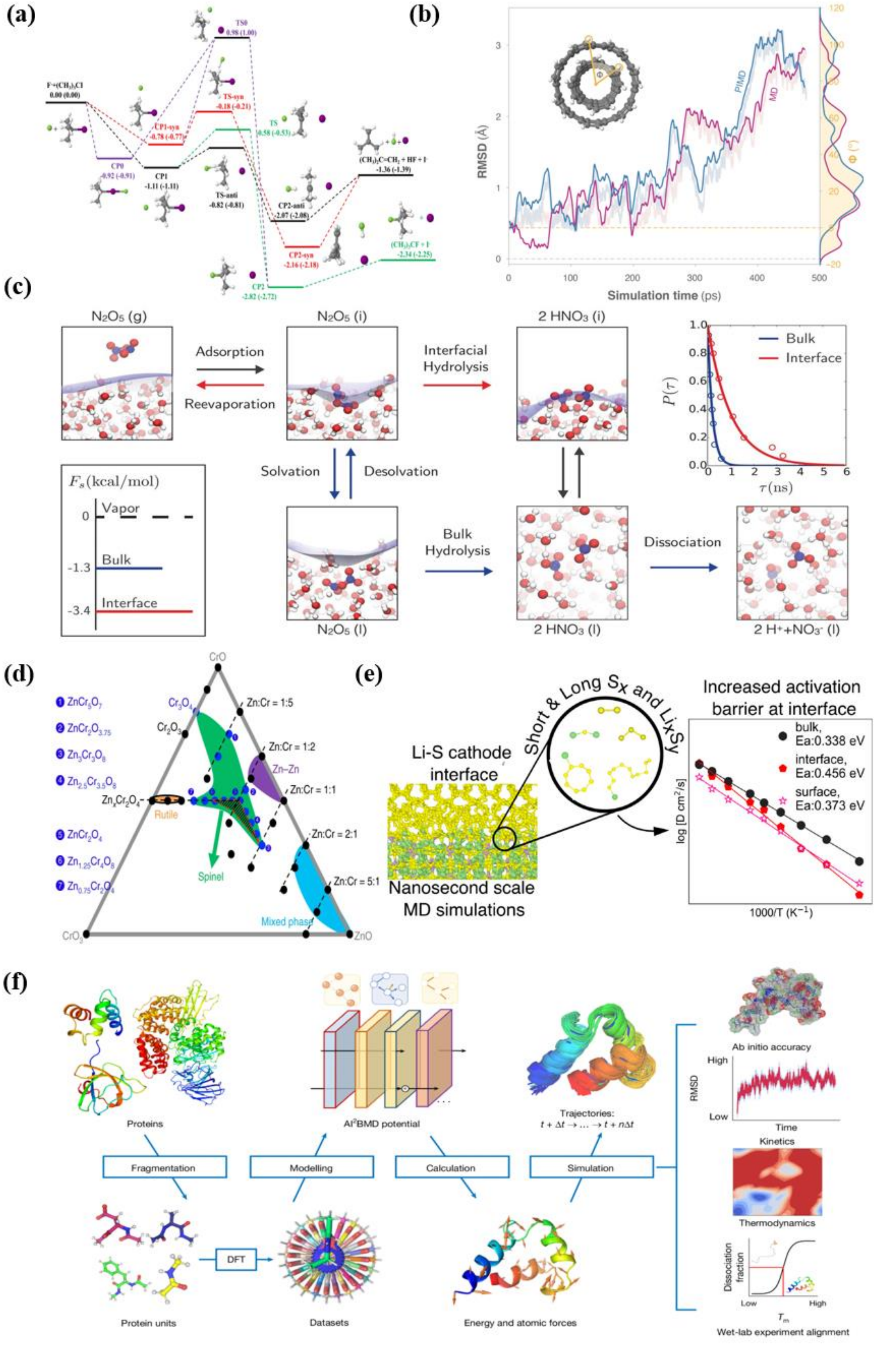
(a)
(b)
Simulation time (ps)
RMSD (Å)
PIMD
MD
(c)
N2O5 (g)
N2O5 (i)
2 HNO3 (i)
Adsorption
Reevaporation
Interfacial Hydrolysis
Bulk
Interface
Solvation
Desolvation
Fs(kcal/mol)
Vapor
Bulk
Interface
Bulk Hydrolysis
Dissociation
N2O5 (l)
2 HNO3 (l)
2 H++NO3- (l)
(d)
ZnCr5O7
ZnCr2O3.75
Zn3Cr3O8
Zn2.5Cr3.5O8
ZnCr2O4
Zn1.25Cr4O8
Zn0.75Cr2O4
Zn:Cr = 1:5
Zn:Cr = 1:2
Zn:Cr = 1:1
Zn:Cr = 2:1
Zn:Cr = 5:1
Rutile
Spinel
Mixed phase
CrO
CrO2
ZnO
(e)
Li-S cathode interface
Short & Long Sx and LixSy
Nanosecond scale MD simulations
Increased activation barrier at interface
bulk, Ea:0.338 eV
interface, Ea:0.456 eV
surface, Ea:0.373 eV
1000/T (K−1)
(f)
Proteins
AI2BMD potential
Trajectories:
Ab initio accuracy
Kinetics
Thermodynamics
Fragmentation
Modelling
Calculation
Simulation
DFT
Protein units
Datasets
Energy and atomic forces
Wet-lab experiment alignment

**Figure 7**. Schematic diagram for selected representative applications in diverse systems, including (a) A FI-NN PES showing multiple channels of the $F^-$ + $(CH_3)_3Cl$ reaction[116], reproduced with permission under the CC-BY 4.0 license, (b) Geometric fluctuation differences as a function of the MD and PIMD simulation time based the sGDML potential of a double-walled nanotube[126], reproduced with permission under the Creative Commons CC-BY-NC license, (c) Elementary physical and chemical steps involved in the reactive uptake of $N_2O_5$ in a pure-water droplet, revealed by a DP model[233], reproduced with permission from Ref. [233] (d) Ternary Zn-Cr-O phase diagram based on the SSW-NN method[237], reproduced with permission from Ref. [237] (e) Li-ion diffusion at the cathode-electrolyte interface with MD simulations using a MTP potential[251], reproduced with permission under the CC-BY 4.0 license and (f) Overall pipeline of the VisNet-based $AI^2BMD$ for protein MD simulations[259], reproduced with permission under the CC BY-NC-ND 4.0 international license.

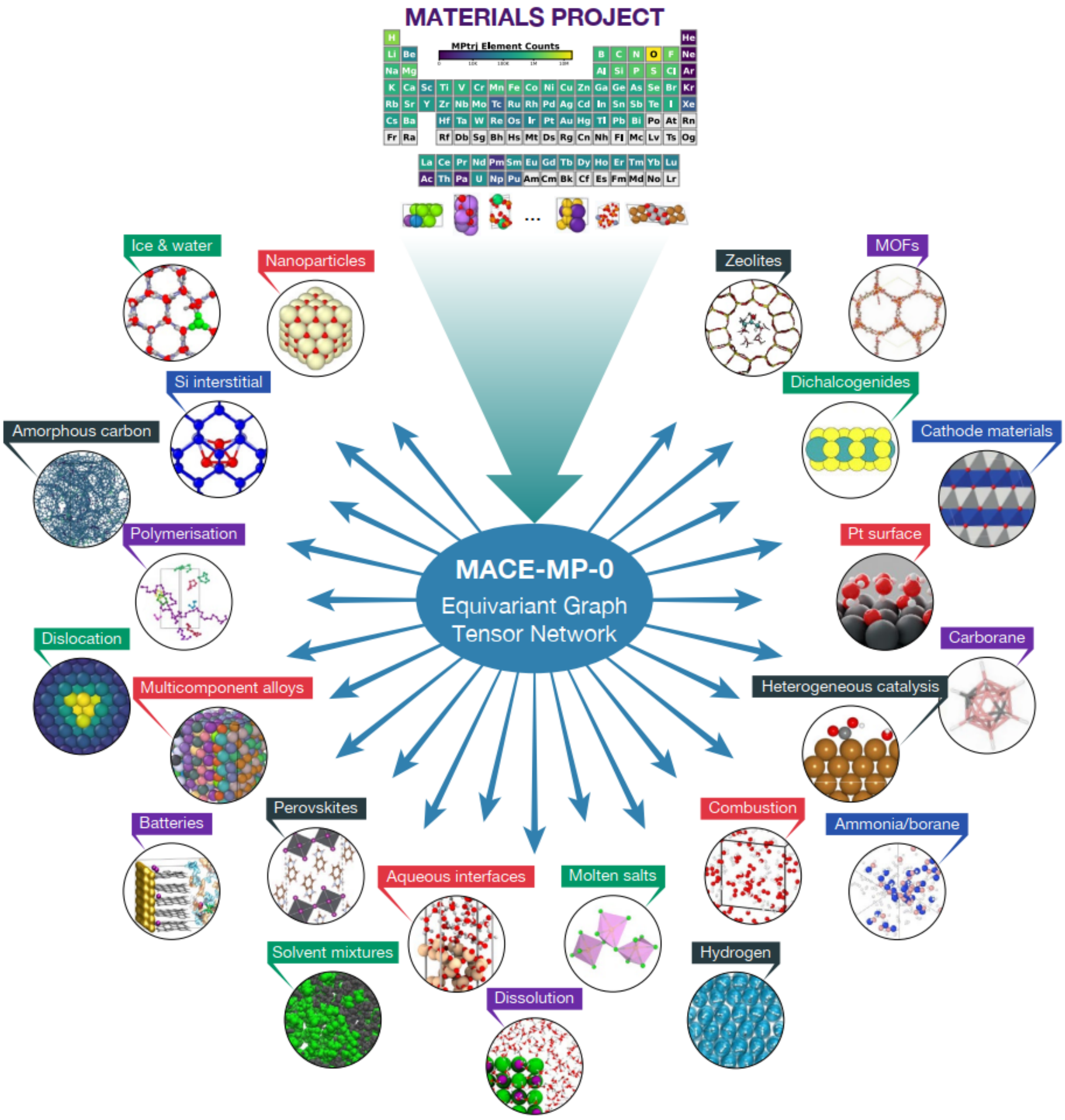

**Figure 8**. A universal potential model, MACE-MP-0[265], that is capable of molecular dynamics simulation across a wide variety of chemistries in the gaseous, solid, liquid phases and interfaces. Reproduced with permission from Ref. 265 under the CC BY-NC-ND 4.0 international license.

**Table I**. A summary of the mainstream MLP programs in alphabetical order and their download links.

| **Software** | **Download link** |
|---|---|
| ænet[137] | http://ann.atomistic.net |
| ACEpotentials.jl[293] | https://github.com/ACEsuit/ACEfit.jl |
| AIMNet[162] | https://github.com/aiqm/aimnet |
| AIMNet2[294] | https://github.com/isayevlab/AIMNet2 |
| ALIGNN[295]/ALIGNN-FF[271] | https://github.com/usnistgov/alignn |
| Allegro[168] | https://github.com/mir-group/allegro |
| AMP[296] | https://bitbucket.org/andrewpeterson/amp/src/master/ |
| ANI-1[132]/ANI-1ccx[280] | https://github.com/aiqm/torchani |
| sGDML[86]/BIGDML[127] | https://github.com/stefanch/sGDML |
| BOTNet[169] | https://github.com/gncs/botnet |
| CACE[184] | https://github.com/BingqingCheng/cace |
| CAMP[186] | https://github.com/wengroup/camp |
| CHGNet[263] | https://github.com/CederGroupHub/chgnet |
| CHIPS-FF[297] | https://github.com/usnistgov/chipsff |
| Cybertron[298] | https://gitee.com/helloyesterday/cybertron |
| Cormorant[160] | https://github.com/risilab/cormorant |
| DeepChem[299] | https://github.com/deepchem/deepchem |
| DP[300]/ DPA-1[274]/ DPA-2[276] | https://github.com/deepmodeling/deepmd-kit |
| DimeNet[159]/DimeNet++[301] | https://github.com/klicperajo/dimenet |
| DTNN[88] | https://github.com/atomistic-machine-learning/dtnn |
| EANN[142] | https://github.com/bjiangch/EANN |
| EDDP[302] | https://github.com/zhubonan/EDDPotentials.jl |
| Equiformer[172] | https://github.com/atomicarchitects/equiformer |
| EquiformerV2[173] | https://github.com/atomicarchitects/equiformer_v2 |
| EquiREANN[187] | https://github.com/bjiangch/EquiREANN |
| FCHL19[139] | https://github.com/qmlcode/qml |
| FitSNAP[303] | https://github.com/FitSNAP/FitSNAP |
| GemNet[167] | https://github.com/TUM-DAML/gemnet_tf |
| Geoformer[304] | https://github.com/microsoft/AI2BMD/tree/Geoformer |
| GNoME[266] | https://github.com/google-deepmind/materials_discovery |
| GPTFF[277] | https://github.com/atomlymaterials-research-lab/GPTFF |
| HermNet[305] | https://github.com/sakuraiiiii/HermNet |
| HotPP[185] | https://gitlab.com/bigd4/hotpp |
| KLIFF[306] | https://github.com/openkim/kliff |
| LASP[211] | http://www.lasphub.com/ |
| M3GNet[262] | https://github.com/materialsvirtuallab/m3gnet |
| MACE[174] | https://github.com/ACEsuit/mace |
| MACE-MP-0[265] | https://github.com/ACEsuit/mace-mp |
| maml[307] | https://github.com/materialsvirtuallab/maml |
| MatterSim[267] | https://github.com/microsoft/mattersim |
| MB-PIPNet[308] | https://github.com/qiyuchem/MB-PIPNet |
| MEGNet[261], | https://github.com/materialsvirtuallab/megnet |
| MindChemistry | https://gitee.com/mindspore/mindscience/tree/master/MindChemistry |
| ML4Chem[309] | https://github.com/muammar/ml4chem |
| MLatom[23]/ ANI-1ccx-gelu[281] | https://github.com/dralgroup/mlatom |
| MLIP[310] | https://gitlab.com/ashapeev/mlip-2 |
| MOLPIPx[105] | https://github.com/ChemAI-Lab/molpipx |
| MSA 2.0[80] | https://github.com/szquchen/MSA-2.0 |
| n2p2[138] | https://github.com/CompPhysVienna/n2p2 |
| NEP[279] | https://gpumd.org/ |
| NequIP[166] | https://github.com/mir-group/nequip |
| NewtonNet[181] | https://github.com/THGLab/NewtonNet |
| OpenChem[311] | https://github.com/Mariewelt/OpenChem |
| Orb[312] | https://github.com/orbital-materials/orb-models |
| PACE[313] | https://github.com/ICAMS/python-pace |
| PANNA[314] | https://gitlab.com/PANNAdevs/panna |
| PES-learn[315] | https://github.com/CCQC/PES-Learn |
| PESPIP[104] | https://github.com/PaulLHouston/PESPIP |
| PFP[273] | https://matlantis.com/ |
| PhysNet[157] | https://github.com/MMunibas/PhysNet |
| PROPhet[316] | https://github.com/biklooost/PROPhet |
| PyXtal_FF[317] | https://github.com/qzhu2017/PyXtal_FF |
| QML-lightning[318] | https://github.com/nickjbrowning/qml-lightning |
| QUIP-GAP[84] | https://github.com/libAtoms/QUIP |
| RANN[319] | https://ranndip.github.io/project_page/ |
| REANN[163] | https://github.com/bjiangch/REANN |
| Runner[136] | https://theochemgoettingen.gitlab.io/RuNNer/ |
| SchNet[156] | https://github.com/atomistic-machine-learning/SchNet |
| SchNetPack2[320] | https://github.com/atomistic-machine-learning/schnetpack |
| SE(3)-Transformers[321] | https://github.com/FabianFuchsML/se3-transformer-public |
| SEGNN[322] | https://github.com/RobDHess/Steerable-E3-GNN |
| SevenNet[323] | https://github.com/MDIL-SNU/SevenNet |
| sGDML[126] | https://github.com/stefanch/sGDML |
| SIMPLE-NN [324] | https://github.com/MDIL-SNU/SIMPLE-NN_v2 |
| SO3KRATES[190] | https://doi.org/10.5281/zenodo.11473653 |
| SpookyNet[161] | https://github.com/OUnke/SpookyNet |
| TeaNet[183] | https://codeocean.com/capsule/4358608/tree/v1 |
| TensorMol[135] | https://github.com/jparkhill/TensorMol |
| torchMD-Net[182] | https://github.com/torchmd/torchmd-net |
| TurboGAP[325] | https://github.com/mcaroba/turbogap |
| ViSNet[171] | https://github.com/microsoft/AI2BMD/tree/ViSNet |

**Table II**. A summary of representative universal potentials and their training details, including the model architecture, model size, training dataset and number of chemical elements covered. Model Size and data size refer to the number of model parameters and data used in the pre-training process. "-" means that the details are not provided in the original paper. Bold numbers correspond to the largest values.

| **Model** | **Architecture** | **Model Size (Million)** | **Training Set** | **Data Size (Million) /Elements** |
|---|---|---|---|---|
| MEGNet[261] | MEGNet[261] | **-** | MP[260] | 0.06/89 |
| M3GNet[262] | M3GNet[262] | 0.2 | MPF[262] | 0.19/89 |
| M3GNet-DIRECT[264] | M3GNet[262] | 1.1 | MPF[262], DIRECT[264] | 1.31/89 |
| CHGNet[263] | CHGNet[263] | 0.4 | MP-trj[263] | 1.58/89 |
| MACE-MP-0[265] | MACE[174] | 4.7 | MP-trj[263] | 1.58/89 |
| GNoME[266] | NequIP[166] | 16.2 | MP[260], OQMD[326], WBM[327] | -/89 |
| MatterSim[267] | M3GNet[262]& Graphormer[328] | **182** | - | 17/89 |
| eqV2-L[269] | EquiformerV2[173] | 153.8 | OMat24[269] | **118**/89 |
| ALIGNN-FF[271] | ALIGNN[295] | 4 | JARVIS-DFT[272] | 0.31/89 |
| PFP[273] | TeaNet[183] | - | PFP[273]* | 42/**96** |
| DPA-1[274] | DP[242] | - | OC2M[275] | 2/56 |
| DPA-2[276] | DPA-2[276] | 7.7 | OC2M[275], Alloy[276], … | 5.1/73 |
| GPTFF[277] | GPTFF[277] | 0.5 | Atomly[277] | 37.8/48 |
| UNEP-v1[278] | NEP[279] | 0.07 | UNEP-v1[279] | 0.1/16 |

*Only a part of the PFP dataset known as HME21[273] is publicly available.